\documentclass[useAMS,usenatbib]{mn2e}

\usepackage{enumitem}
\usepackage{calc}
\usepackage{chngcntr}
\usepackage{epsfig}
\usepackage{amsmath}
\usepackage{natbib}
\usepackage{graphicx}
\usepackage[caption=false]{subfig}
\usepackage{wrapfig}
\usepackage[T1]{fontenc}
\usepackage{enumitem}
\usepackage{hyperref}
\usepackage[normalem]{ulem}
\usepackage{textcomp}
\usepackage{longtable}
\usepackage{answers}
\usepackage{array}
\usepackage{tablefootnote}
\title[Data Reduction Comparison]
  {The JCMT Gould Belt Survey: A Quantitative Comparison Between SCUBA-2 Data Reduction Methods}
\author[S. Mairs et al]
  {S. Mairs$^{1, 2}$, D. Johnstone$^{1, 2, 3}$, H. Kirk$^{2}$, S. Graves$^{3, 4}$, J. Buckle$^{5, 6}$, \newauthor S.F. Beaulieu$^{7}$, D.S. Berry$^{3, 4}$,  H. Broekhoven-Fiene$^{1, 2}$, M.J. Currie$^{3, 4}$, \newauthor M. Fich$^{7}$, J. Hatchell$^{8}$, T. Jenness$^{3, 9}$,  J.C. Mottram$^{10}$, D. Nutter$^{11}$, K. Pattle$^{12}$, \newauthor J.E. Pineda$^{13}$, C. Salji$^{5, 6}$,  J. Di Francesco$^{2}$, M.R. Hogerheijde$^{10}$, D. Ward-Thompson$^{12}$, \newauthor and the JCMT Gould Belt survey team
\\
$^{1}$Department of Physics and Astronomy, University of Victoria, Victoria, BC, V8P 1A1, Canada\\
$^{2}$NRC Herzberg Astronomy and Astrophysics, 5071 West Saanich Rd, Victoria, BC, V9E 2E7, Canada\\
$^{3}$Joint Astronomy Centre, 660 North A`oh\={o}ok\={u} Place, University Park, Hilo, Hawaii 96720, USA\\
$^{4}$East Asian Observatory, 660 North A`oh\={o}ok\={u} Place, University Park, Hilo, Hawaii 96720, USA\\
$^{5}$Astrophysics Group, Cavendish Laboratory, J J Thomson Avenue, Cambridge, CB3 0HE, UK\\
$^{6}$Kavli Institute for Cosmology, Institute of Astronomy, University of Cambridge, Madingley Road, Cambridge, CB3 0HA, UK\\
$^{7}$Department of Physics and Astronomy, University of Waterloo, Waterloo, Ontario, N2L 3G1, Canada Ê\\
$^{8}$Physics and Astronomy, University of Exeter, Stocker Road, Exeter EX4 4QL, UK\\
$^{9}$Large Synoptic Survey Telescope Project Office, 933 N. Cherry Ave, Tucson, Arizona 85721, USA\\
$^{10}$Leiden Observatory, Leiden University, PO Box 9513, 2300 RA Leiden, The Netherlands\\
$^{11}$School of Physics and Astronomy, Cardiff University, The Parade, Cardiff, CF24 3AA, UK\\
$^{12}$Jeremiah Horrocks Institute, University of Central Lancashire, Preston, Lancashire, PR1 2HE, UK\\
$^{13}$Institute for Astronomy, ETH Zurich, Wolfgang-Pauli-Strasse 27, CH-8093 Zurich, Switzerland}

\date{Released 2015 Xxxxx XX}

\pagerange{\pageref{firstpage}--\pageref{lastpage}} \pubyear{2015}

\def\LaTeX{L\kern-.36em\raise.3ex\hbox{a}\kern-.15em
    T\kern-.1667em\lower.7ex\hbox{E}\kern-.125emX}

\begin{document}

\label{firstpage}

\maketitle

\clearpage

\begin{abstract}

Performing ground-based submillimetre observations is a difficult task as the measurements are subject to absorption and emission from water vapour in the Earth's atmosphere and time variation in weather and instrument stability. Removing these features and other artifacts from the data is a vital process which affects the characteristics of the recovered astronomical structure we seek to study. In this paper, we explore two data reduction methods for data taken with the Submillimetre Common-User Bolometer Array-2 (SCUBA-2) at the James Clerk Maxwell Telescope (JCMT). The JCMT Legacy Reduction 1 (JCMT LR1) and The Gould Belt Legacy Survey Legacy Release 1 (GBS LR1) reduction both use the same software (\textit{Starlink}) but differ in their choice of data reduction parameters. We find that the JCMT LR1 reduction is suitable for determining whether or not compact emission is present in a given region and the GBS LR1 reduction is tuned in a robust way to uncover more extended emission, which better serves more in-depth physical analyses of star-forming regions. Using the GBS LR1 method, we find that compact sources are recovered well, even at a peak brightness of only 3 times the noise, whereas the reconstruction of larger objects requires much care when drawing boundaries around the expected astronomical signal in the data reduction process. Incorrect boundaries can lead to false structure identification or it can cause structure to be missed. In the JCMT LR1 reduction, the extent of the true structure of objects larger than a point source is never fully recovered.

\end{abstract}


\section{Introduction}
\label{introductionsec}

In an effort to probe the cold dust in several well-known nearby star-forming regions with the James Clerk Maxwell Telescope (JCMT), the Gould Belt Survey (GBS, \citealt{wardthompson2007}) has been performing submillimetre continuum observations using the Submillimetre Common-User Bolometer Array 2 (SCUBA-2). The SCUBA-2 instrument \citep{holland2013} is a wide field submillimetre bolometer camera with an unprecedented 10,000 pixels. The device maps regions at $450\mathrm{\:} \mu$m and $850\mathrm{\:}\mu$m simultaneously with effective resolutions of 9.6$\arcsec$ and 14.1$\arcsec$ \citep{dempsey2013}, respectively. Mapping the sky over 100 times faster than its predecessor, SCUBA \citep{holland1999}, this next generation detector has offered star formation researchers the chance to analyse nascent stellar systems at submillimetre wavelengths with a higher efficiency and a broader context than ever before with a single dish, ground-based telescope (see, for example, \citealt{sadavoy2013}, \citealt{salji2015}, and \citealt{pattle2015}). 

Reducing SCUBA-2 data is a complex process with a variety of solutions, each designed to best uncover particular features (e.g. bright, compact emission versus faint, diffuse emission). This variety is, in part, due to the nature of the observations. While performing observations using the PONG scanning mode (the method employed in this study; see \citealt{kackley2010} and \citealt{holland2013}), the JCMT continuously scans back and forth across the sky at differing angles to fill in a circular pattern. By visiting the same locations from different angles and at different times, sources of low frequency drift are manifested as a length-scale feature which can be separated from the sources of interest so that spatially invariant structures can be identified. These sources of drift include instrument-based noise (see \citealt{holland2013}) but are dominated by atmospheric noise, which varies temporally. In this way, many individual bolometers will observe each sky location. Constructing a final image requires identifying significant structure and removing those large-scale features created by the noise. Quantifying how the observing process affects the observed signal is therefore of utmost importance as it characterises how confident we can be that we are detecting true astronomical emission and not artificial constructs at each scale of interest. 

Determining the optimal image reconstruction of submillimetre bolometer data has been an area of acute interest for researchers using JCMT data since before the operation of the original SCUBA instrument. For example, \cite{richer1992} presented the ``maximum entropy'' reconstruction method, which is based on the assumption that the most likely reconstruction is the one which has maximum entropy (see \citealt{narayan1986}, for a review). \cite{wilsonmaxentropy} and \cite{pierce-price2002} went on to develop maximum entropy reconstruction methods for data taken by SCUBA. Several other methods were also employed. For instance, \citet{borys2002} wrote a SCUBA map-making algorithm based on a least squares approach while \cite{johnstoneinversion} developed a ``matrix inversion'' reconstruction which accounted for the varying uncertainty in a given map's background noise. The latter was based on a technique for the Wilkinson Microwave Anisotropy Probe (WMAP) described by \cite{wright1996}. At approximately the same time, \cite{jenness2000} implemented a Fourier deconvolution method developed by \cite{emerson1995} (Emerson 2) as the reduction method for the standard SCUBA data analysis software at the JCMT.

The issue of effective and accurate submillimetre data reduction is not unique to the JCMT. For example, the Bolocam Galactic Plane Survey (BGPS; see \citealt{aguirre2011} for the 1.1mm survey and \citealt{merello2015} for the 350 $\mu$m images) uses an algorithm called \textit{CRUSH} (The Comprehensive Reduction Utility for SHARC-II; \citealt{kovacs2006}) to reduce the 350 $\mu$m continuum images taken using the Submillimetre High Angular Resolution Camera II (SHARC-II; \citealt{dowell2003}) at the Caltech Submillimetre Observatory (CSO). Like SCUBA-2, SHARC-II is a ground-based bolometer array which must contend with instrumental imperfections, atmospheric interference, and electronic noise. The CRUSH algorithm is an iterative process which determines a series of gain and weight estimations for each array component one-by-one while filtering noise spikes and other bad data (such as cosmic rays and electronic discharges; see \citealt{kovacs2006} for more information). 
Other examples include \textit{SANEPIC}, the maximum-likelihood map-making algorithm used for the Balloon-borne Large Aperture Submillimeter Telescope (BLAST; \citealt{patanchon2008}) experiment which uses a series of approximations to reduce the required computational power necessary for image production, and the AzTEC (Aztronomical Thermal Emission Camera, located at the Large Millimetre Telescope) Data Reduction Pipeline, which uses Principal Component Analysis (PCA) to model and remove correlated components of the bolometer signals. The latter method, however, is only ideal for the recovery of compact structures as extended emission will have correlated components that will be flagged as noise \citep{scott2008}. 

In this paper, we explore the data reduction method used by the JCMT GBS team for their first Legacy Release (GBS LR1; Kirk et al., in prep.) of SCUBA-2 data by first comparing and contrasting with another method employed by the JCMT science and computing group (JCMT LR1, Graves et al., in prep.) then analysing how well each method preserves structures with known properties. We focus here only on the 850 $\mu$m maps. The same arguments can be extended to 450 $\mu$m maps but there is less high signal-to-noise data available for comparison and the absolute calibration uncertainties are higher. 
 
To construct an image from the raw SCUBA-2 data, the GBS LR1 and the JCMT LR1 reduction teams both use the \emph{makemap} algorithm found in Starlink's\footnote{\url{http://www.starlink.ac.uk}} {\emph{smurf}} package (\citealt{chapin2013}; also see the SCUBA-2 data reduction cookbook \citealt{sc21}). The \emph{makemap} algorithm employs an iterative technique which applies flat field corrections, performs an extinction correction,  models and removes noise correlated between detectors, estimates and masks the astronomical signal, and measures the noise of each bolometer contributing to each pixel in the reduced image. In total, for all the different observational strategies available, \emph{makemap} has over one hundred configurable parameters which are supplied in a text file called the ``Dynamic Iterative Map Maker configuration file'', or, ``\emph{dimmconfig} file''. With these parameters, the user has significant control over the entire reduction algorithm from the preprocessing stages to what is included in each model (astronomical, noise, etc.). Varying the parameters supplied in \emph{makemap}'s \emph{dimmconfig} file, therefore, will give rise to different final images. Many of these differences will be subtle but any differences between reductions should be well understood. 


The \emph{dimmconfig} file employed is dependent on the scientific goals desired. For example, since the beginning of the SCUBA-2 mapping initiative, the JCMT Science Archive's \citep{economou2015} Legacy Release project has been focused on producing public images and subsequent catalogues which could answer the simple question: ``Is there compact emission in an observed region?''.  In this way, the archive can provide a useful set of information to a wide variety of user projects without doing more complex analyses that require careful human oversight. 

To this end, the JCMT LR1 data reduction was tuned to downgrade extended structure while effectively identifying compact emission.  The GBS SCUBA-2 team, however, has a wide range of specific scientific interests including but not limited to an investigation of the emissivity spectral index  of the thermal dust emission (\citealt{hatchell2013}, Coud\'e et al., in prep.) and its relationship to temperature and column density (in conjunction with Herschel Space Telescope data; \citealt{sadavoy2013}, Chen et al., in prep.), structure mapping (\citealt{salji2015}, \citealt{salji2015cores}), investigations into protoplanetary disks (\citealt{dodds2015}, \citealt{buckle2015}, Broekhoven-Fiene et al., in prep.), fragmentation analyses (Mairs et al., in prep.), virial studies \citep{pattle2015}, clustering (Lane et al., in prep., Kirk et al., in prep.), and radiative feedback processes around young stars \citep{rumble2015}. The GBS LR1 is the most current data reduction product available from the GBS team to suit these individual goals while remaining consistent across all the star-forming regions observed by the GBS. This broad mandate means that the data reduction strategy requires both compact and extended emission recovery. The latter is a more difficult process as the separation of large, faint structures from time-varying noise is non-trivial. To this end, the GBS LR1 reduction employs a two-step process in which significant recovered emission is first identified automatically based on the signal-to-noise ratio (SNR) then, based on those resulting maps, a rigorous analysis is performed wherein the observer uses the structure detected and in some cases knowledge from other data sets (for example, from the Herschel Space Observatory) to define a boundary around any emission that is likely real. After the significant structure is more rigorously identified by the observer, the map undergoes a second round of data reduction.

The format of this paper is as follows: In Section \ref{dimmconfigsec}, we outline in more detail the specific differences between the JCMT LR1 reduction and GBS LR1 reduction methods. In Section \ref{comparisonsec}, we compare the final maps produced using each method in three regions of interest within the Orion A South star-forming complex. In Section \ref{completenesssec}, we explore the scales and flux levels at which the reductions preserve structure by constructing artificial Gaussian sources and recovering them after they have been processed by \emph{makemap} in a pure noise field. In Section \ref{changemasks}, we investigate the effect of changing the flux density threshold at which SCUBA-2 data are considered to be significant astronomical signal. We then carry on this investigation by determining the effects of changing the size of a user-defined boundary around emission that has been labelled as significant. In Section \ref{observablessec}, we give a brief overview of how data reduction can affect the results of common observational metrics used when studying star forming regions. Finally, in Section \ref{conclusionsec}, we summarise the main results and present our conclusions.

\section{Data Reduction Parameters}
\label{dimmconfigsec}

The SCUBA-2 data reduction process, \emph{makemap}, is explained in detail by \citet{chapin2013} (see their Figure 6).  To summarise here, the raw data from each scan is first assembled into a continuous time series and a flat field correction is applied. Then, the data is downsampled and discontinuities such as steps, spikes, and gaps are repaired. Following that, the mean of each bolometer time series is removed from the map and the iterative portion of the mapmaking procedure begins.    

There are six models iteratively constructed and improved upon in five steps before the final map is created:

\begin{enumerate}[labelindent=0pt,labelwidth=\widthof{\ref{last-item}},label=\arabic*.,itemindent=1em,leftmargin=!]

\item The COM and GAI models first estimate and remove the common mode signal across the bolometers at each time step, respectively. The common mode removal can be performed over the full SCUBA-2 focal plane array ($\sim$400$\arcsec$ scale) or over each sub-array individually ($\sim$200$\arcsec$ scale; see the \texttt{com.perarray} parameter in Sections \ref{JSAreductionsec} and \ref{GBSreductionsec}, below). 

\vspace{3mm}

\item The EXT model then corrects for extinction.

\vspace{3mm}

\item Next, the FLT model (based on a high-pass filtering algorithm) removes independent low-frequency noise associated with each individual bolometer directly from the time series. The physical scale to which we filter is initially input as a length (see the \texttt{flt.filt\_edge\_largescale} parameter in Sections \ref{JSAreductionsec} and \ref{GBSreductionsec}, below) since the scanning speed of the telescope is known and thus a length scale can be converted to a time scale. 

\vspace{3mm}

\item The AST model then identifies significant astronomical signal in the estimated map and removes its projection from the time series so that the noise can be measured accurately. 

\vspace{3mm}

\item Finally, the NOI model measures the noise in the residual time series after the removal of the AST signal.

\end{enumerate}

If the solution converges, the algorithm produces the final map. Otherwise, the process repeats itself by inverting the previous solution and re-estimating each model until convergence is achieved.

All continuum observations presented in this paper are taken from the GBS survey \citep{wardthompson2007} and were made with the PONG1800 mapping mode \citep{bintley2014} which produces individual circular regions called ``pongs'' with a usable diameter of $\sim$0.5\textdegree. To map large regions, as in the case of Orion A South (see Section \ref{comparisonsec}), circular pongs are placed so their edges overlap to allow for a more uniform noise level in the final mosaic. This final mosaicked dataset includes four to six repeats of every $\sim$0.5\textdegree\ observing field, with a higher number of repeats for data taken in worse weather conditions. The maps are created with 3$\arcsec$ pixels and the final images, originally in units of picowatts (pW), are converted to Jy Beam$^{-1}$ using the peak intensity conversion factor $537\: \mathrm{Jy}\: \mathrm{pW}^{-1}\: \mathrm{Beam}^{-1}$ \citep{dempsey2013} for the GBS LR1 reduction and $567\: \mathrm{Jy}\: \mathrm{pW}^{-1}\: \mathrm{Beam}^{-1}$ (Graves et al. in prep) for the JCMT LR1 reduction. For the GBS LR1 reduction, the iterative process was terminated when the average pixel value changed by less than 0.1\% of the estimated map rms. In contrast, the JCMT LR1 reduction uses a termination value of 1\%. The difference in the stopping criteria (and other paramaters) was based on the specific goals of each data reduction. A more stringent threshold was used for the GBS LR1 because this allows more diffuse, extended structure to be recovered. In contrast, the less stringent threshold was sufficient for the JCMT LR1 reduction as it was only important to reach convergence after recovering bright, compact structure which takes less computing time (see Sections \ref{JSAreductionsec} and \ref{GBSreductionsec}, below). Note that the three regions presented here have a noise level of $\sigma_{rms} \approx 0.0038 \mathrm{\:Jy\:Beam}^{-1}$ (for a description and analysis of the entire Orion A South region, see Mairs et al. in prep.). In the following two subsections, we discuss individual choices for a subset of \emph{makemap}'s configurable parameters for each of the JCMT LR1 reduction and GBS LR1 reduction methods.

 
\subsection{JCMT LR1 Data Reduction}
\label{JSAreductionsec}

The JCMT science and computing group is currently producing for release to the wider astronomy community a uniform reduction and co-addition of its publicly available SCUBA-2 850 $\mu$m data (for more information, see \cite{bell2014} and Graves et al. in prep.). This legacy release will consist of a) individual Hierarchical Equal Area isoLatitude Pixelization (HEALPix) tiles with equal area but some non-square pixels (see \citealt{gorski2005}) reduced using a map-maker configuration chosen to work for a diversity of regions and observational types; b) these individual observations coadded together to produce HEALPix tiles covering all the regions observed; and c) catalogues of the emission detected in each map.

The \emph{makemap} parameters chosen are available with the Starlink {\emph{smurf}} map-maker software in a text file named \textit{dimmconfig\_jsa\_generic.lis}. The parameters were developed with a focus on minimising the possibility of artificial emission being created during the reduction process while still producing high quality results across a diversity of observation types towards a range of astronomical regions. To accomplish this task, it was decided that no attempt to recover large-scale structure would be made. In addition, given the automated nature of these reductions, external masks rigorously defined by an observer (like in the case of the GBS LR1, below) were not a viable option, and so, a more restrictive automasking configuration was used. The following are the most important of the \emph{makemap} parameters:

\begin{description}

\item[\texttt{com.perarray = 1}] This parameter creates a separate common mode for each sub-array of SCUBA-2, which means that sources larger than the sub-array size ($\sim$200$\arcsec$) will not be recovered. As stated previously, accurately recovering compact structure is one of the main goals of the JCMT LR1 reduction; this reduction is not suitable for an analysis of extended structure.

\vspace{2mm}

\item[\texttt{flt.filt\_edge\_largescale = 200}] This parameter filters all emission on scales above 200$\arcsec$ and is consistent with \texttt{com.perarray=1}; it operates in the time stream and uses the scanning speed to convert the length scale to a time scale.

\vspace{2mm}

\item[\texttt{numiter = -25} \& \texttt{ast.skip = 5}] The first 5 iterations are done without an AST model and then up to a further 20 iterations are performed. The reduction will exit at that point even if it hasn't converged. Processing of all the Orion observations, however, converged. This choice of 25 iterations helps keep the reduction of a very large number of observations to a reasonable time scale.

\vspace{2mm}

\item[\texttt{ast.zero\_snr = 5}] Pixels which have a value of at least 5 $\sigma_{rms}$ will be identified as astronomical signal.

\vspace{2mm}

\item[\texttt{ast.zero\_snrlo = 3}] This parameter allows identified sources with pixel values of at least 5 $\sigma_{rms}$ to expand in area until the flux density values are 3 $\sigma_{rms}$.

\vspace{2mm}

\item[\texttt{maptol = 0.01}] This parameter specifies when to terminate the mapmaking procedure. Using this reduction method, the process will terminate when the average pixel value in the map changes by less than 1\% of the estimated map RMS.



\end{description}

\subsection{The Gould Belt Legacy Survey Legacy Release 1}
\label{GBSreductionsec}

The GBS LR1 reduction is a two-part process. In part one, a reduction similar in approach to the JCMT LR1 reduction is run, i.e., flux is assigned to the AST model based on pixels with high signal-to-noise-ratios. This is referred to as the ``automask'' reduction.

Part two is an additional step which is employed in the GBS LR1, but not the JCMT LR1 reduction. After the automask reduction is performed, the individual maps are coadded for a higher SNR and the resulting image is used to define regions of likely emission. The boundaries drawn around the significant signal become the user-defined ``external mask'' and this mask is used to perform a second round of data reduction to better recover faint and extended structure. In this re-reduction, instead of basing the AST model on pixels which achieve a specific SNR value, the pixels within the external mask boundaries are defined as containing astronomical signal. This allows for a more well-defined masked area around structure which we are confident is real to be included in the AST model when compared using a single observation.

Major differences from the GBS Internal Release 1 reduction method (IR1; see \citealt{hatchell2013}, \citealt{buckle2015}, \citealt{rumble2015}, and \citealt{pattle2015}) include smaller pixel sizes (GBS LR1 pixels are half as wide as GBS IR1 pixels, allowing for better characterization of small-scale emission as well as more accurate peaks and positions of compact sources), and additional filtering of the raw data which better prevents the appearance of large-scale noise features in the reduced maps. Note that despite the attempts to minimize noise and reconstruct the diffuse emission, there are still challenges in retrieving all the signal for large sources (see Section \ref{completenesssec}). The following are the most important parameters to be compared with the JCMT LR1 reduction, above. Unless otherwise stated, these parameters are the same for both the automask reduction and the external mask reduction:

\begin{description}

\item[\texttt{com.perarray = 0}] No separate common mode for each sub-array of SCUBA-2 is created. Thus, sources with sizes up to the full array size ($\sim$400$\arcsec$) can be confidently recovered. Sources with sizes approaching and exceeding this value will have large-scale features subtracted by the common mode model.

\vspace{2mm}

\item[\texttt{flt.filt\_edge\_largescale = 600}] This parameter, operating on the time stream, filters all emission on scales above 600$\arcsec$. This value is three times the size of the JCMT LR1 reduction emission scale. 

\vspace{2mm}

\item[\texttt{flt.filt edge\_largescale\_last = 200}] Only on the last iteration, the emission outside the AST mask is filtered at 200$\arcsec$ instead of 600$\arcsec$. This parameter was defined to help suppress uncertain emission structure outside the masked regions. The detection of extended emission and the accuracy of our calibration are trusted within the masked regions but outside the mask boundaries we cannot be confident that the extended structure is real. By filtering the unmasked data harshly on the final iteration, we remove the large-scale features and uncover any underlying, robust, small-scale sources. 

\vspace{2mm}

\item[\texttt{numiter = -300} \& \texttt{ast.skip = 5}] The first five iterations are done without an AST model, and up to a further 295 iterations are performed. The reduction will exit at that point even if it hasn't converged. All processing of the Orion observations, however, converged.

\vspace{2mm}

\item[\texttt{ast.zero\_snr = 5}] Pixels which have a value of at least 5 $\sigma_{rms}$ will be identified as astronomical signal. This only applies to the automask reduction as the external mask reduction defines what will be considered as astronomical signal using clear physical boundaries.

\vspace{2mm}

\item[\texttt{ast.zero\_snrlo = 0}] This parameter does not allow sources identified as astronomical signal  to extend to lower flux densities. Instead, a pixel must have a value of at least 5 $\sigma_{rms}$ to be included in the AST mask in the automask reduction. While the final reduction employs a user-defined external mask, the boundaries of that external mask will be based on the sources identified by the automask.
 
\vspace{2mm}

\item[\texttt{maptol = 0.001}] This parameter specifies when to terminate the mapmaking procedure. Using this reduction method, the process will terminate when the average pixel value in the map changes by less than 0.1\% of the estimated map RMS. 
 
\end{description}

\section{Data Reduction Comparison in Orion A South}
\label{comparisonsec}

\begin{figure*}
\centering
\includegraphics[width=16cm, height=15cm]{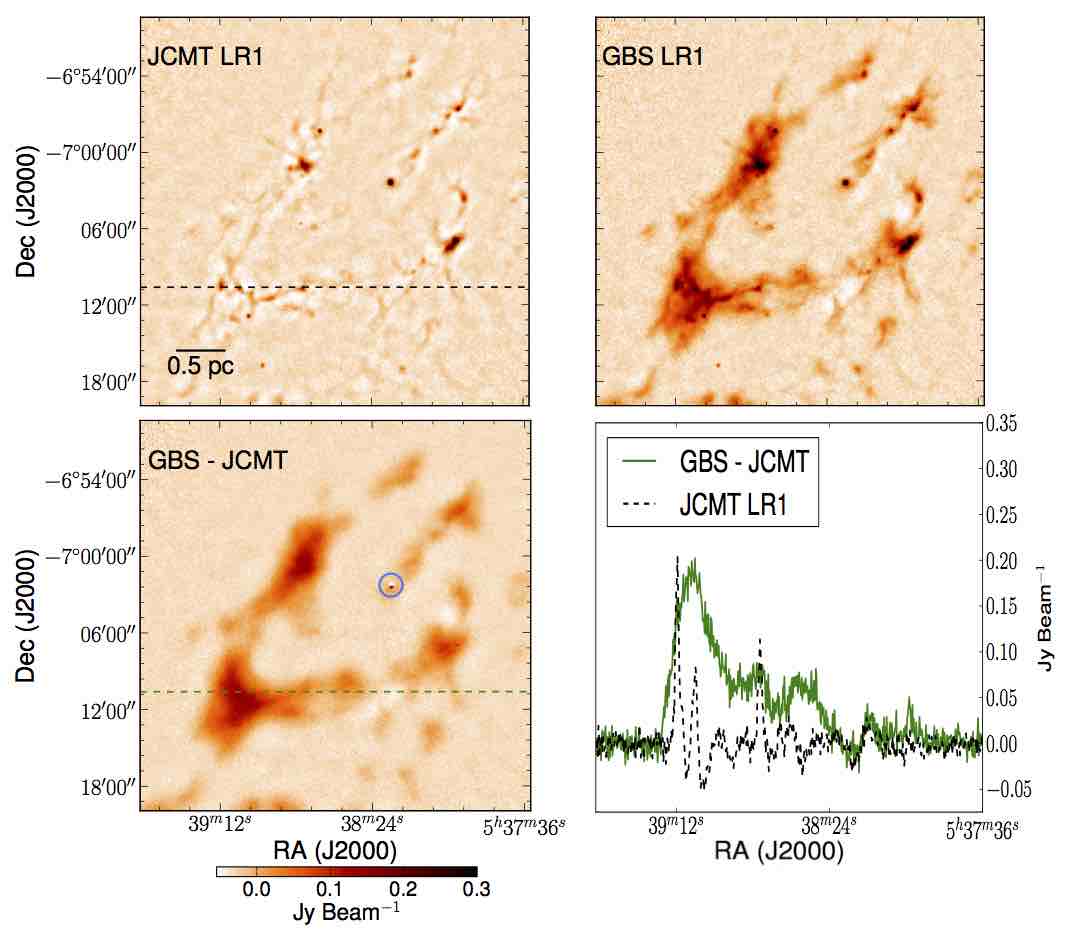}
\caption{Region 1: The first of three representative regions of Orion A South.  \emph{Top left:} 850 $\mu$m SCUBA-2 image reduced with the JCMT LR1 reduction parameters. \emph{Top right:} 850 $\mu$m SCUBA-2 image reduced with the GBS LR1 reduction parameters including the external mask. \emph{Bottom left:} The GBS LR1 map subtracted by the JCMT LR1 map. The blue circle indicates a peculiarity in the map due to realigning the HEALPix projection of the JCMT LR1 reduction to the tangent plane projection of the GBS LR1 reduction. \emph{Bottom right:} The intensities of the JCMT LR1 map subtracted from those of the GBS LR1 map across the positions corresponding to the dotted line shown in the bottom left (green) and the intensities of the JCMT LR1 map across the same coordinates (black).}
\label{orionasouthcompareregion0}
\end{figure*}
\begin{figure*}
\centering
\includegraphics[width=16cm,height=15cm]{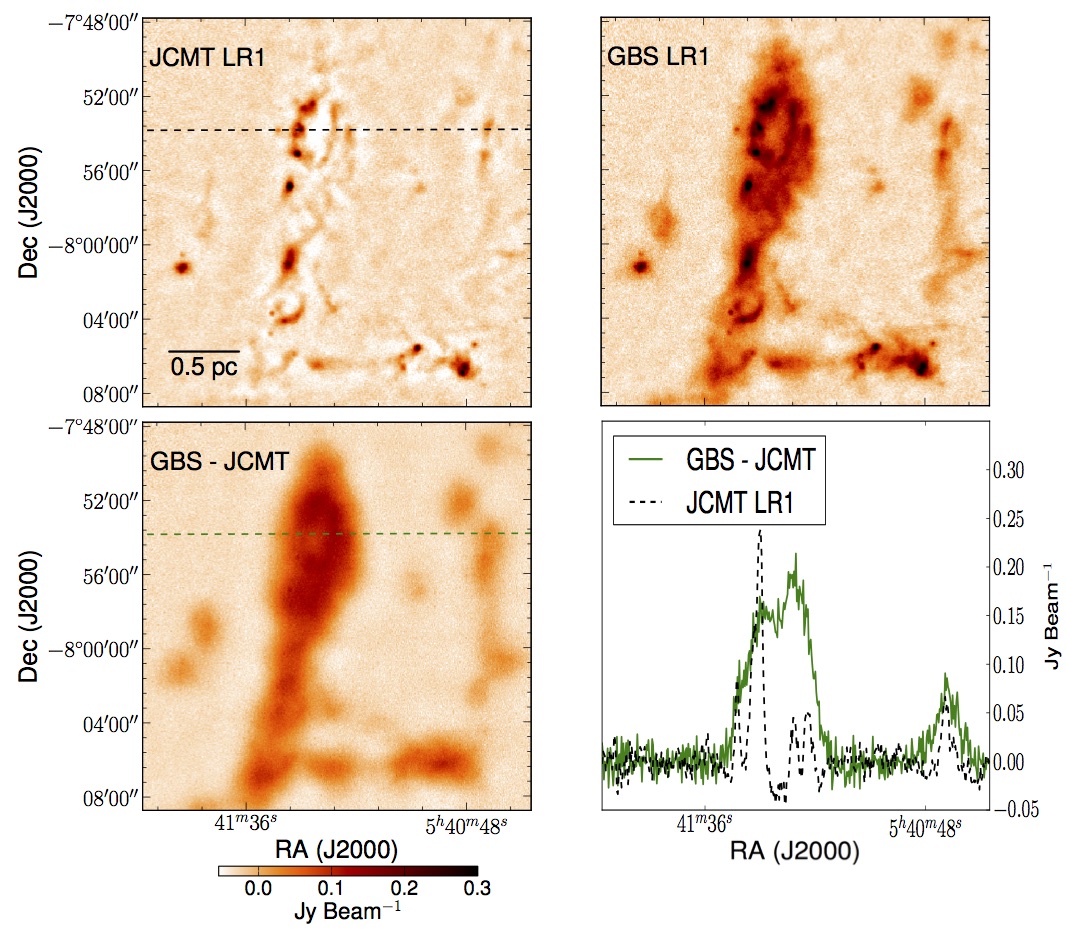}
\caption{Region 2: The second of three representative regions of Orion A South. Each panel is presented in the same manner as in Figure \ref{orionasouthcompareregion0}.}
\label{orionasouthcompareregion1}
\end{figure*}
\begin{figure*}
\centering
\includegraphics[width=16cm,height=15cm]{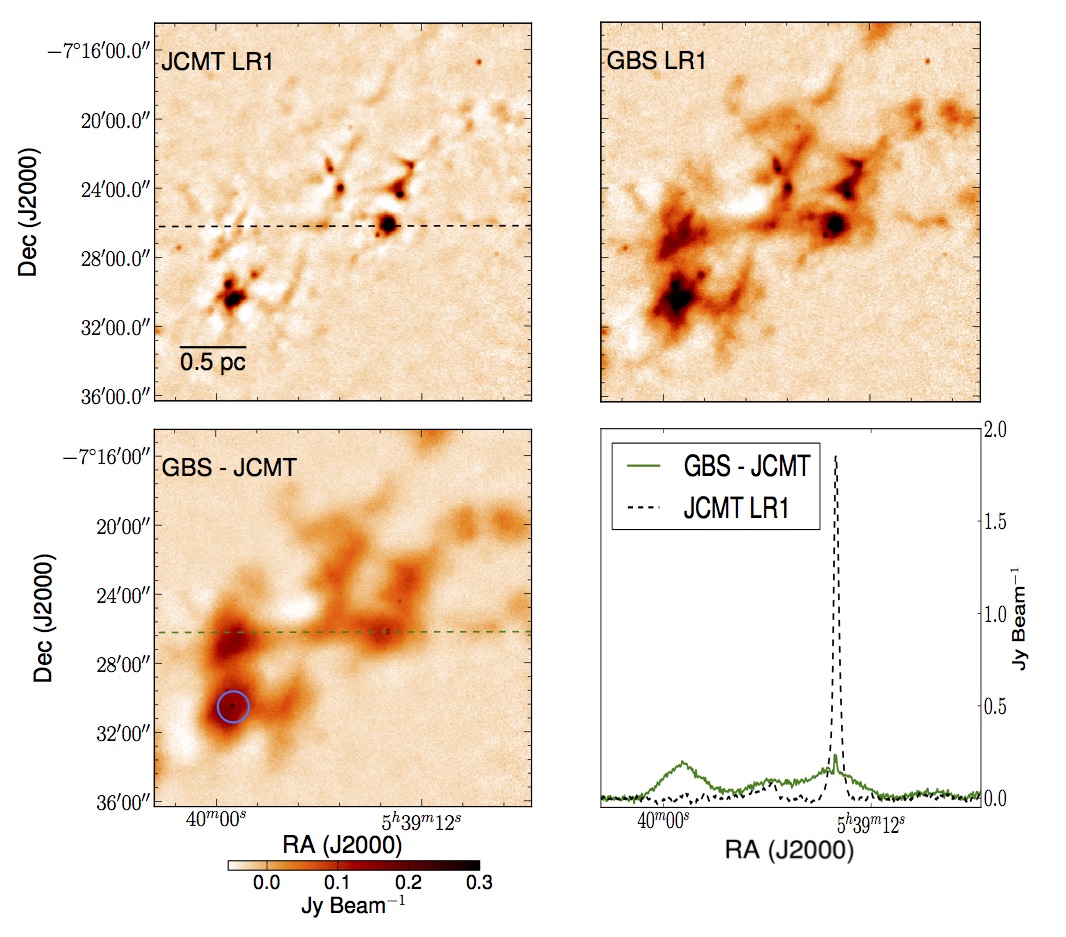}
\caption{Region 3: The final representative region of Orion A South discussed. Each panel is presented in the same manner as in Figure \ref{orionasouthcompareregion0}. The blue circle indicates an example of a residual peak left over after the subtraction.}
\label{orionasouthcompareregion2}
\end{figure*}

In Figures \ref{orionasouthcompareregion0} to \ref{orionasouthcompareregion2}, we compare two reductions using representative areas of the GBS-defined ``Orion A South'' region by resampling the JCMT LR1 HEALPix images to match the GBS LR1 pixel size and projection. Orion A South is a 2.2\textdegree \hspace{0.3mm} $\times$ 3.1\textdegree \hspace{0.3mm} subsection of the Orion cloud complex, an active star-formation site approximately 450 pc from the Sun (see \citealt{johnstone2006}, \citealt{allen2008}, and \citealt{davis2009} for more information). The entire Orion cloud complex has a mass in excess of $2 \mathrm{\:x\:} 10^{5} M_{\odot}$ \citep{wilson2005} and is comprised of two individual molecular clouds: Orion A and Orion B (\citealt{reipurth2008}; page 459: Overview of the Orion Complex; \citealt{megeath2012}, and references therein). The northern section of the Orion A Cloud (Orion A North) is home to the Orion Nebula and the well-known integral shaped filament \citep{johnstone1999}. The Orion A South region is also a target of interest and, although less complex and dense, it exhibits several different stages of star formation (see Mairs et al. in prep. for more information on Orion A South).


From Figures \ref{orionasouthcompareregion0} to \ref{orionasouthcompareregion2}, we see that the qualitative difference between the JCMT LR1 and the GBS LR1 reductions rests in the extended emission. As expected, compact structures present in the data are well accentuated in the former. Indeed, if one's goal is simply to determine where compact emission exists in the map, this style of data reduction is entirely reasonable. One must, however, be cautious when analysing the JCMT LR1 maps any further as the total flux density present in a given region will likely be underestimated even in tight boundaries around the brightest sources (see Section \ref{differencequant} below). 

The GBS reconstruction gives a more accurate picture of the large-scale structure, as required for the science goals of the consortium. Beyond simply identifying where emission exists, the intent here is to recover the full emission structure. The less drastic filtering parameter and the manual external masking process allow, for example, much more extended emission associated with a given object to be included in stability calculations, providing the opportunity to characterise the wide varieties of dense gas/dust morphologies seen in all regions of the survey. In comparison, the JCMT LR1 reduction's recovery of fractional extended emission inhibits these goals. 
 
In the bottom left panels of Figures \ref{orionasouthcompareregion0} to \ref{orionasouthcompareregion2}, we present maps depicting the JCMT LR1 reduction subtracted from the GBS LR1 legacy reduction for each representative region of Orion A South. With only a few exceptions (blue circles have been drawn around prominent examples in Figures \ref{orionasouthcompareregion0} and \ref{orionasouthcompareregion2}), we see that the compact structure has been almost completely canceled out and only the extended emission remains. This situation is not unexpected since both reduction techniques process compact objects in the same manner. Instead, the significant differences are found in the large-scale emission. 

To see the similarity of the compact structure between the two reduction techniques more easily, the bottom right panel of each Figure shows the flux density through a slice of each region. The solid green line shows the GBS LR1 intensities with the JCMT LR1 intensities subtracted while the dashed black line shows only the JCMT LR1 intensities across the same cut. The physical positions of each slice is represented in the top left and bottom left panels of each figure by the dashed lines. Indeed, the features seen are easily recognisable by comparing the image to the intensity plots. We see in many cases that the peaks in the JCMT LR1 image are fainter than the peaks in the GBS LR1 image. This is because the GBS LR1 data reduction recovers more of the underlying flux density when using an external mask covering a larger area than the automatically detected emission in the JCMT LR1 map. Evidently, the better recovery of this extended structure occasionally increases the flux density of small scale features associated with the parent source, resulting in features like that which is indicated by the blue circle in Figure \ref{orionasouthcompareregion2}. The residual peak intensity left over in this example is 12\% of the peak intensity measured in the GBS LR1 image. The amount in which the flux density increases depends on the properties of the larger-scale parent source. This difference can cause discrepancies where the JCMT LR1 peak intensities are as low as 40\% of the GBS LR1 peak intensities in the observed regions of Orion A South. In most cases, however, the JCMT LR1 peak intensities are $\sim$60\% of the GBS LR1 peak intensities (see Figure \ref{orionasouthscatterplot}). The brightest peaks have the smallest differences. Note that fainter, compact, isolated peaks also appear to have a bias to slightly lower intensities in the JCMT LR1 map due to the harsh filtering parameter and less thorough masking procedure. This effect, however, is not as strong as the pedestal caused by the recovery of extended structure.


Compact objects near the boundaries of HEALPix tiles can deform and elongate. The blue circle in the bottom left panel of Figure \ref{orionasouthcompareregion0} highlights a peculiarity resulting from converting the HEALPix projection employed by the JCMT LR1 reduction to the tangent plane projection used by the GBS LR1. The oblong shape of the circled source could potentially alter the position of a detected peak in this region. We note, however, that this event is uncommon and we only see this one example in the entire Orion A South map.

\subsection{Quantitative Differences between the GBS LR1 and JCMT LR1 reductions}

\subsubsection{Structure Identification Algorithm}

To compare the two reductions fairly, we need to identify structure in a consistent manner across each of the two maps. This situation is where the GBS LR1's diffuse, extended emission presents us with a challenge. There are many structure identification algorithms freely available to apply to data such as Orion A South and each program will break up large-scale structures or amalgamate smaller ones in different ways (see \citealt{stutzki1990}; \citealt{williams1994}; \citealt{rosolowsky2008}; \citealt{getsources2012}; and \citealt{berry2015} for algorithm examples). We note that several individual JCMT LR1 ``sources'' can reside within a single GBS LR1 reduction ``source'' no matter which algorithm is used to identify structure, due to the lack of extended emission in the former defining smaller areas of significant signal.

\begin{figure*}
\centering
\includegraphics[width=17.5cm,height=9.5cm]{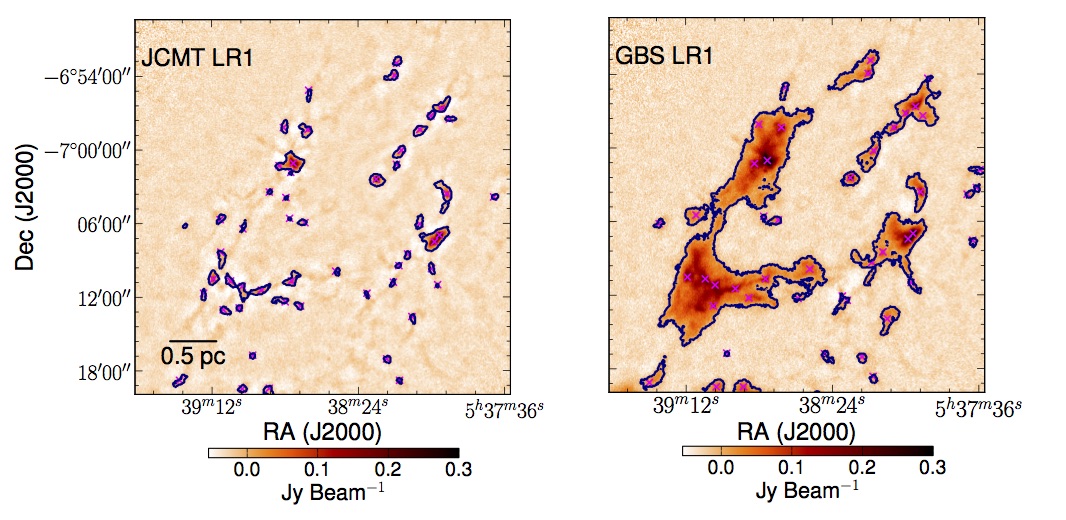}
\caption{Islands (blue contours) and peaks (magenta Xs) identified by the \emph{FellWalker} algorithm for both reductions tested using Region 1 as an example. See text for information on the basic algorithm parameters used to identify structure.}
\label{islandspeaksexample}
\end{figure*}


Since we are specifically interested in the differences between compact structure and extended structure separately, for this analysis we employ the JCMT Science Archive algorithm \emph{jsa\_catalogue} found in \textit{Starlink's} \emph{PICARD} package \citep{sun265}. The catalogues generated with this algorithm are based around the concept of \emph{islands} (or extents) and \emph{peaks} (blue contours and magenta Xs, respectively in Figure \ref{islandspeaksexample}). The routine was designed to do a good job of cataloging all regions where emission was strongly detected (the islands), and then to provide a list of local peak positions within each island to categorise the nature of the emission. Source catalogues based on this approach will be released along with the JCMT LR1 maps  (Graves et al., in prep.).

The \emph{jsa\_catalogue} routine was used to identify regions of contiguous emission above a minimum number of pixels (9 pixels), using an SNR cut of 5 (the noise level in these maps is $\sim0.0038 \mathrm{\:Jy\: Beam}^{-1}$). These regions are found using the \emph{FellWalker} algorithm \citep{berry2015} as implemented in the {\emph{cupid}} package \citep{berry2007} on the SNR map, and then their shapes are extracted from the data map so that the total flux density, average noise across the region, and the flux density at the peak can be calculated. These regions are identified as the islands or extents of emission.

Within each of these detected regions, the routine then searches for peaks, again using the \emph{FellWalker} algorithm. The routine is configured to identify a pixel as a peak if it has a) a peak value of magnitude 5 times higher than the average noise in the island; b) a minimum dip of magnitude 5 times grater than the average noise towards the nearest larger peak; and c) more pixels than the minimum of 9 assigned towards this peak (to avoid noise clumps). The peaks are identified only by their position and brightness. No attempt is performed to assign a shape to them. Selecting this set of parameters was a result of extensive testing of the \emph{FellWalker} algorithm on many different SCUBA-2 maps as well as maps which only included artificial structure. There is no standard set of parameters as the algorithm can be tuned differently depending on the specific scientific question of interest. More generally, the debate about which boundaries to draw around different structures, especially in regions with complex emission, has been present in the literature for many years (see, for example, \citealt{pineda2009}) and it should be performed carefully and with an approach that addresses specific goals.

\subsubsection{Properties of the Structure in Each Reduced Map}
\label{differencequant}

\begin{table*}
\caption{Summary of the three regions analysed and the number of sources found within each. The RA and DEC represent the centre coordinates of each region.}
\label{orionasouthsourcenumtable}
\begin{tabular}{|c|c|c|c|c|c|c|}
\hline
 Region & RA, DEC & Size &\multicolumn{2}{c|}{Number of Peaks}&\multicolumn{2}{c|}{Number of Islands}\\
  & & & GBS&JCMT&GBS&JCMT\\
\hline\hline
1 &  5h38m35.632s, -7:04:39.80 & 0.51$^{\circ}$ $\times$ 0.51$^{\circ}$ & 47 & 51 & 25 & 48  \\
2 &  5h41m16.139s, -7:58:16.84 & 0.35$^{\circ}$ $\times$ 0.35$^{\circ}$&35 & 35 & 13 & 32\\
3 &  5h39m30.365s, -7:25:23.70 & 0.37$^{\circ}$ $\times$ 0.37$^{\circ}$ &29 & 44 & 12 & 35\\
\hline
\end{tabular}
\end{table*}


For both reductions, Table \ref{orionasouthsourcenumtable} shows the number of peaks and the number of islands detected in each region shown in Figures \ref{orionasouthcompareregion0} to \ref{orionasouthcompareregion2} along with the centre coordinates and sizes of each region. Table \ref{orionasouthcomparetable} provides the total, maximum, and median effective radius and total flux density found in each population of island sources as well as the peak intensity found in each population of sources identified by \emph{jsa\_catalogue} in each reduced dataset. The effective radius given is the radius of a circle with the same area as the object of interest.

\begin{table*}
\caption{Comparison of the identified structure in the GBS LR1 and JCMT LR1 reductions. Three metrics are used to compare the GBS LR1 and JCMT LR1 methods in the three regions. The areas are calculated by summing the number of pixels within a given source identified by \emph{jsa\_catalogue's} island catalogue, the total flux densities are the summation of the pixel values in each source's footprint, and the peak intensities refer to the sources identified by \emph{jsa\_catalogue's} peak catalogue.}
\label{orionasouthcomparetable}
\begin{tabular}{|l|c|c|c|c|c|c|}
\hline
\multicolumn{1}{|c|}{Region: Metric} & \multicolumn{2}{c|}{Total} & \multicolumn{2}{c|}{Maximum} & \multicolumn{2}{c|}{Median}\\
  & GBS& JCMT& GBS&JCMT&GBS&JCMT\\
\hline\hline
1: Area (arcsec$^{2}$) & 3.8$\times$10$^{6}$ & 6.1$\times$10$^{5}$ & 2.5$\times$10$^{6}$ & 7.4$\times$10$^{4}$ & 1.4$\times$10$^{4}$ & 7.2$\times$10$^{3}$  \\
1: Total Flux Density (Jy) &  142.4 & 17.5  & 101.8 & 4.0 & 0.3 & 0.1  \\
1: Peak Intensity (Jy Beam$^{-1}$) &   -- & -- & 1.3 & 1.0 & 0.1 & 0.1 \\\hline\hline
2: Area (arcsec$^{2}$) &  3.3$\times$10$^{6}$ & 5.2$\times$10$^{5}$ & 2.4$\times$10$^{6}$ & 1.1$\times$10$^{5}$ & 5.5$\times$10$^{4}$ & 6.2$\times$10$^{3}$   \\
2: Total Flux Density (Jy) & 149.0 & 18.4  & 118.4 & 4.4 & 2.2 & 0.1   \\
2: Peak Intensity (Jy Beam$^{-1}$) &  -- & -- & 0.8 & 0.6 & 0.2 & 0.1  \\\hline\hline
3: Area (arcsec$^{2}$) &  3.1$\times$10$^{6}$ & 6.3$\times$10$^{5}$ & 2.7$\times$10$^{6}$ & 1.4$\times$10$^{5}$ & 3.3$\times$10$^{4}$ & 6.4$\times$10$^{3}$  \\
3: Total Flux Density (Jy) &  127.7 & 26.9  & 119.2 & 10.0 & 0.8 & 0.1   \\
3: Peak Intensity (Jy Beam$^{-1}$) & -- & -- & 2.6 & 2.4 & 0.2 & 0.1\\
\hline
\end{tabular}
\end{table*}

\begin{figure*}
\centering
\subfloat{\label{gaussincosfieldfig}\includegraphics[width=9.7cm,height=9cm]{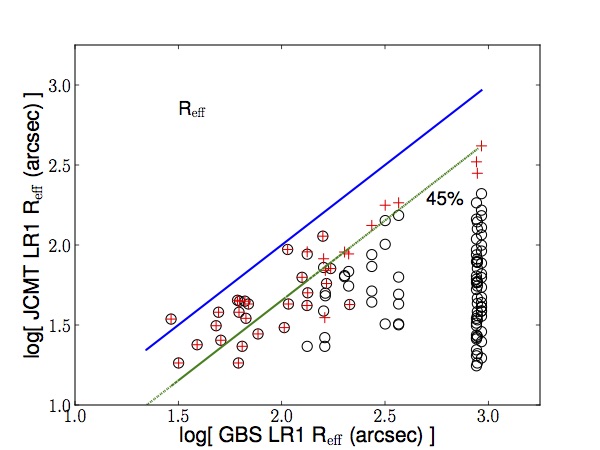}}
\subfloat{\label{noisefieldalonefig}\includegraphics[width=9.7cm,height=9cm]{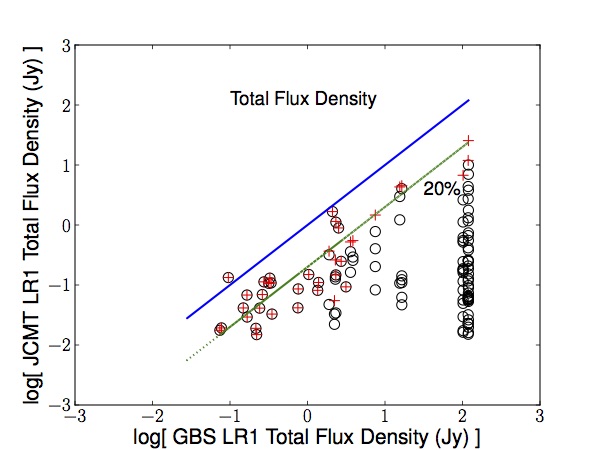}}\\
\subfloat{\label{gaussincosfieldfig}\includegraphics[width=9.7cm,height=9cm]{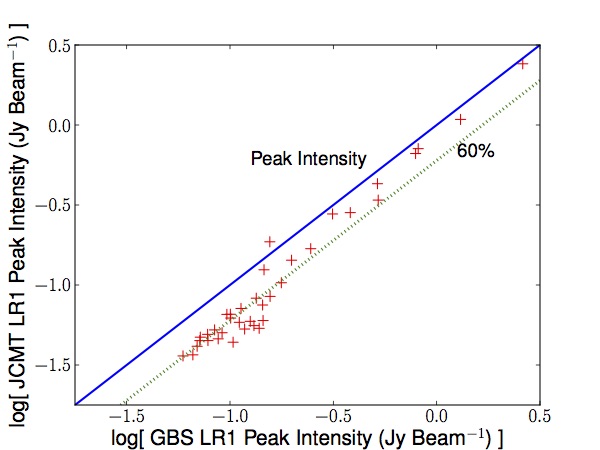}}
\caption{Comparing R$_{\mathrm{eff}}$ (the radius of a circle with the same area as a given source identified by \emph{jsa\_catalogue's} island catalogue), total flux density, and peak intensity between the two data reductions in the three representative regions of Orion A South. In the former two, all the JCMT LR1 reduction islands associated with a given GBS LR1 island are plotted in black. We sum the effective radius and the total flux density of all of the associated JCMT LR1 islands, respectively, and plot the total as a red plus sign. For the peak intensities, we plot the peak sources with the maximum flux density identified by \emph{jsa\_catalogue's} peak catalogue within a given associated island. The blue (solid) lines show a one to one relationship and the green (dotted) lines have a slope of unity at the shown percentage of the GBS LR1 values.}
\label{orionasouthscatterplot}
\end{figure*}

Figure \ref{orionasouthscatterplot} shows a comparison of the source radii, total flux densities, and peak intensities identified in all three regions. Since the islands we identify in the JCMT LR1 reduction are smaller and more fragmented than the sources we identify in the GBS LR1 reduction, we match our JCMT LR1 reduction sources with the associated GBS LR1 reduction ``parent source'' so that we can compare specific structures in the same location of each map. We assign JCMT LR1 sources to a GBS LR1 parent based on the position of the geometric centre of a given source. If a JCMT LR1 source centre falls within the boundaries of a GBS LR1 island, we consider those emission structures to be associated.    There are only a few cases where a JCMT LR1 source is not associated with a GBS LR1 source or \emph{vice versa} but these are small, faint objects which have little bearing on our final results. Thus, for the radii and total flux densities, we only include objects in the Figure if they have a counterpart in each image. When comparing the peak locations between the GBS LR1 and JCMT LR1 images, the overwhelming majority of detected peaks lie within one pixel of their counterpart. Notably, there are relatively small numbers of isolated GBS LR1 peaks and JCMT LR1 peaks across the entire Orion A South region, this is simply due to the degree of smoothness of the diffuse emission recovered by each respective reduction (see Section \ref{peaksectionorionasouth}). As Table \ref{orionasouthsourcenumtable} shows, region 3 has the largest disparity in peak number due to the manner in which the bright, diffuse emission is broken up by the JCMT LR1 reduction. Many borderline peaks are identified that would not have been if the underlying continuous structure was more visible. The results presented here only include peaks which are associated with the same emission in each map.

In Figure \ref{orionasouthscatterplot}, the JCMT LR1 reduction source property is on the ordinate and the GBS LR1 reduction source property is on the abscissa. For the radii and total flux densities, we plot each of the JCMT LR1 reduction sources associated with a given GBS LR1 source in black and we sum all of the associated JCMT LR1 sources and plot that as a red plus sign. With this approach we can compare how much emission is being found in each region in a fair way. For the peak intensity, we directly compare \emph{jsa\_catalogue's} catalogue detections found in each map produced by the two data reduction methods. The blue lines in Figure \ref{orionasouthscatterplot} all show a one to one relationship and the green lines have a slope of unity at the shown percentage of the GBS LR1 values. The latter is meant to give an indication of the representative difference between the two reductions based on each measurement.

Since the \emph{jsa\_catalogue} algorithm gives us information on both the peaks as well as the extended structures within a given map, we discuss each of these aspects individually.

\subsubsection{Peaks} 
\label{peaksectionorionasouth}


The GBS LR1 and JCMT LR1 reductions have similar compact structure in regions without extended flux (see, for example, the isolated source to the left of the main structure in the top two panels of Figure \ref{orionasouthcompareregion1}). This similarity is because the main differences in the data reduction parameters deal with structure that is much greater than one 14$\arcsec$ beam. For example, the \texttt{flt.filt\_edge\_largescale} parameter (see Section \ref{dimmconfigsec}) chosen by each reduction impacts recovered emission on the scale of $>$ 200$\arcsec$. As well, the larger number of iterations employed in producing the GBS LR1 maps will not have a large effect on the bright, compact structure, but that parameter is set for the recovery of faint emission seen at larger scales.

Since more extended emission is recovered in the GBS LR1 reduction than in the JCMT LR1 reduction, more faint locations are detected as individual sources. The JCMT LR1 reduction, however, produces maps which divide up large structures into individual, compact components. The overall effect is that the number of peaks detected in the GBS LR1 and the JCMT LR1 maps are very similar. In two of the regions, there are more JCMT LR1 map peaks since locations that contain diffuse emission in the GBS LR1 maps will not meet the peak detection criteria (but do meet, in some cases, the island detection criteria) whereas the more fragmented JCMT LR1 maps will contain borderline detections in these regions (see Table \ref{orionasouthsourcenumtable}). It is clear from the statistics presented in Table \ref{orionasouthcomparetable} that, in general, the peak intensities are consistently higher in the GBS LR1 map in every region. As seen previously (recall the bottom right panel of Figure \ref{orionasouthcompareregion2}), the externally masked GBS reduction changes the amount of flux occasionally on the small-scale as well as the large-scale in regions with extended emission (see Figure \ref{orionasouthscatterplot}). Evidently, going deeper into the GBS LR1 image with an external mask and allowing that structure to grow raises the peaks by a pedestal. The pedestal is not constant, however, because it depends on the larger-scale, fainter emission structure.  This behaviour is made clear once more in Figure \ref{orionasouthscatterplot} which shows a positive correlation between the brightest JCMT LR1 peak intensities and their associated GBS LR1 peak intensities in log-log space. Indeed, most of the brightest JCMT LR1 peak intensities below 1 Jy Beam$^{-1}$ are only $\sim$60\% as bright as their GBS LR1 counterparts because the diffuse background is not included in the former. As the peaks become brighter in each map, they also become more similar to one another. No aperture fitting is performed in this study, however. We perform no background subtraction for the flux extraction to account for large-scale structure. 

\subsubsection{Extended Structure}

We see many similarities between the three regions shown in Figures \ref{orionasouthcompareregion0} to \ref{orionasouthcompareregion2}. In fact, when using \emph{jsa\_catalogue} to investigate the significant structure in the maps, it was clear that the identified sources in each JCMT LR1 image largely trace the same structure as the corresponding GBS LR1 image. The lack of diffuse emission connecting the bright, compact peaks in the JCMT LR1 map, however, causes \emph{jsa\_catalogue} to divide up the larger structures seen in each GBS LR1 map into many smaller ones. Therefore, between the two reduction methods, the number of ``extended'' sources identified in each region is always higher for the JCMT LR1 map (see Table \ref{orionasouthsourcenumtable}). The lack of extended emission in the JCMT LR1 map is shown clearly in the relative area occupied by identified sources in Orion A South. The total flux density of all the identified sources is closely related to the area since the structure identification algorithm identifies roughly the same peaks in both maps. 
 
In general, we see that the JCMT LR1 image sources are smaller, as expected, except for the smallest (sometimes faint and spurious) sources present in both reduced images. A positive correlation is followed in radii in log-log space but it is not one to one. We find that the summed radii of all JCMT LR1 islands detected within one GBS LR1 island are typically $\sim$45\% those of the GBS LR1 island radius (as seen in Figure \ref{orionasouthscatterplot}). This corresponds to a total JCMT LR1 island area of only $\sim$20\% of the GBS LR1 island area which is exactly what we see in the total flux density relationship of Figure \ref{orionasouthscatterplot}.

\section{Completeness Testing}
\label{completenesssec}

\begin{figure}
\hspace{1cm} \includegraphics[width=6cm,height=15cm]{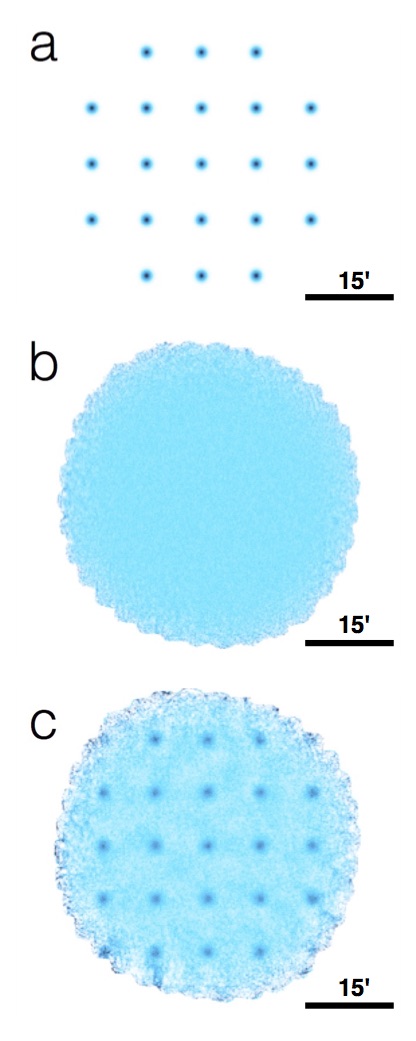}
\caption{\emph{Top, a}: An example Gaussian grid. Here, each Gaussian has a FWHM of 7 beams and a peak of 9 $\sigma_{rms}$. The constructed grids are spaced accordingly for the given Gaussian FWHM. When the Gaussians are smaller, more sources are added to the noise field. \emph{Middle, b}: The field nearly devoid of structure in which the Gaussians were inserted. \emph{Bottom, c}: The final map depicting the 7 beam FWHM, 9 $\sigma_{rms}$ peak Gaussians combined with the noise field using the GBS LR1 reduction method.}
\label{drfig}
\end{figure}

To test how well each data reduction method preserves structure, we study the effect of \emph{makemap} on artificial Gaussians with known properties. We construct and ``observe'' a broad range of Gaussians by varying the sources' full widths at half maximum (FWHM) and peak intensity values. In this section, we approximate 1 beam to be 15$\arcsec$. For each FWHM and peak value, we insert a grid of Gaussian sources into a pure noise field via \emph{makemap}'s \emph{fakemap} parameter (see Figure \ref{drfig}). In this manner, the Gaussian grids are added to the raw time stream of the data and are subjected to the usual data reduction steps. We inserted 32 sets of Gaussians in total with FWHMs of 1 beam (15$\arcsec$), 3 beams (45$\arcsec$), 5 beams (75$\arcsec$), and 7 beams (105$\arcsec$). For each of these FWHM values, we use a series of peak intensity values between 3 $\sigma_{rms}$ and 25 $\sigma_{rms}$.  

The field in which we insert these regular grids of Gaussian sources is an 850 $\mu$m field nearly devoid of structure obtained by the Cosmology Legacy Survey (CLS; \citealt{geach2013}) in the PONG1800 mapping mode (see Figure \ref{drfig}). This field is a 42 minute integration of a circular region with a diameter of $\sim$0.8 degrees observed on September 30$^{th}$, 2013; it is a small subset of the CLS data in this region and, thus, it provides a representative sample of the noise in typical SCUBA-2 images in this configuration. We identify structure in the reduced dataset by employing a Gaussian curve fitter\footnote{See {\emph{SciPy}}'s generalised \emph{curve\_fit} function: \url{docs.scipy.org/doc/scipy-0.14.0/reference/generated/scipy.optimize.curve\_fit.html}} at the known position of the sources located away from the noisy edges of the map. By a simple visual analysis, it appears that artificial structure outside the inserted Gaussians is effectively non-existent. To use the most robust method of returning the true structure, we provide the fiducial Gaussian location, peak intensity, and FWHM values as initial ``guesses'' for the peak coordinates, peak intensity, and the size of the expected structure to the fitting algorithm. We then compare the obtained output properties of each fitted Gaussian to the nominal input properties. 

\begin{figure}%
\centering
\subfloat{\label{externalmaskfig}\includegraphics[width=7cm,height=7cm]{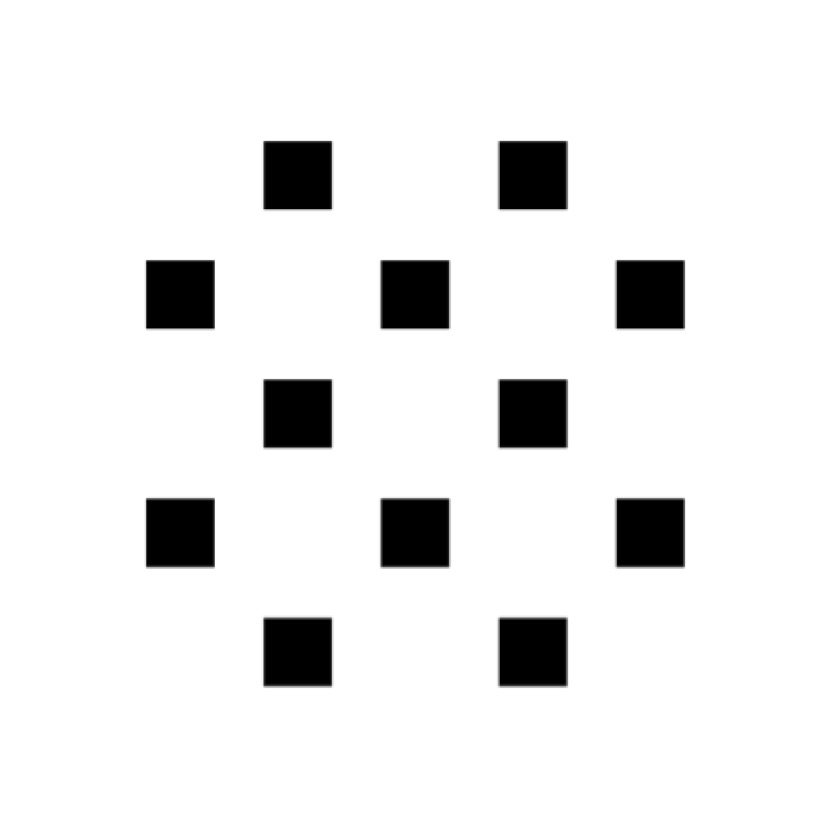}}\\
\subfloat{\label{externalmaskedgaussfig}\includegraphics[width=9cm,height=7cm]{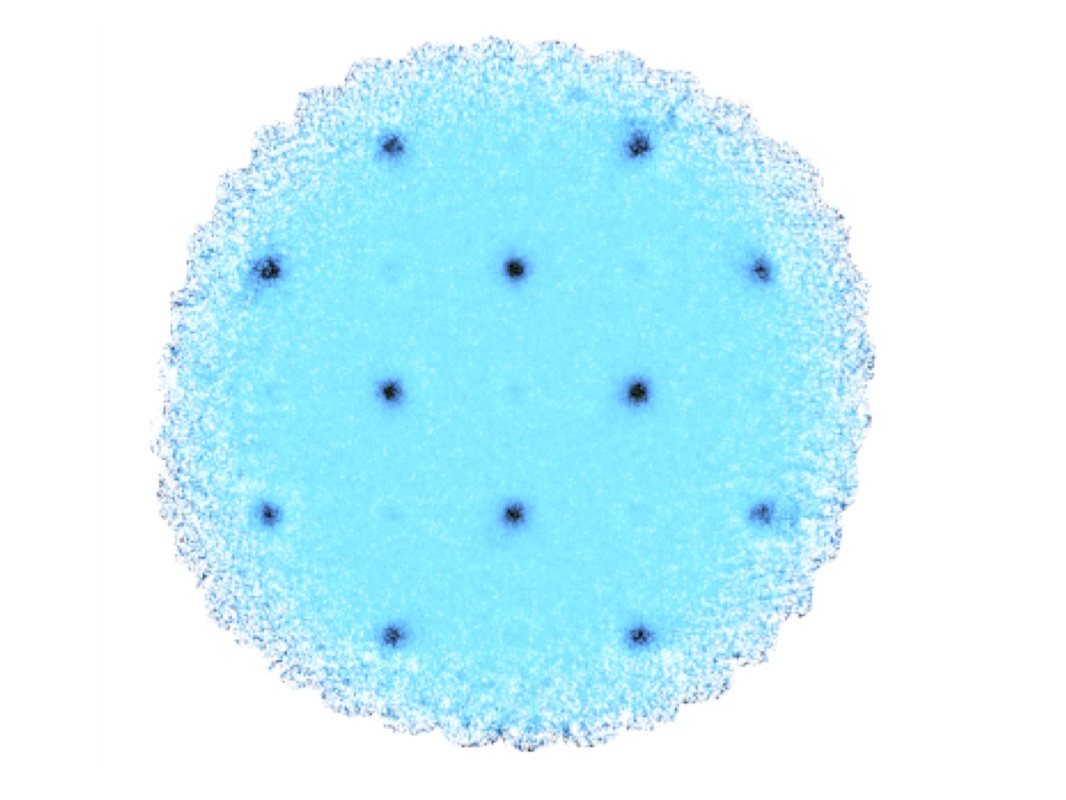}}
\caption{\emph{Top}: The checkerboard pattern of the external mask for the 7 beam FWHM Gaussians. Black indicates the positive mask. \emph{Bottom}: The final map after the GBS LR1 reduction using the checkerboard mask on the 7 beam FWHM, 9$\sigma_{rms}$ peak Gaussians in the noise field.}
\label{extmaskfig}
\end{figure}

We run each Gaussian grid run through \emph{makemap} three times. In the first case, we use the standard GBS LR1 automasking procedure (see Section \ref{changeautomask} for tests of different automask parameters). Recall, however, that the GBS LR1 reduction employs a user-defined external mask based on this type of automask for the final product. The second time, we apply an external mask which covers only half of the Gaussians in the image. The external mask we use is not generated using the normal GBS LR1 external masking procedure. Instead, we use a simple checkerboard pattern of square masks, where each mask is 20 beams (5$\arcmin$) on a side and centred upon every second Gaussian source (see Figure \ref{extmaskfig} for an example and Section \ref{changeexternalmask} for the effect of changing the size of the masks). For the final execution of \emph{makemap}, we reduce the data using the JCMT LR1 reduction parameters. 


With these three sets of maps, we are able to compare the output properties of the GBS LR1 automasked Gaussians, the Gaussians lying inside and outside an external mask, and the JCMT LR1 reduction Gaussians. In addition, we construct a fourth set of maps wherein the artificial Gaussians are spatially added directly to the reduced noise field. We call this latter set the ``non-DR'' case, as \emph{makemap} was not used for Gaussian reconstruction. 

It is pertinent to note that these tests are all based on Gaussian structures. There is, however, a wide variety of clump morphologies, from nearly spherical dense cores to long, thin filaments. Consideration of the effects of data reduction on Gaussians should inform even those more complex sources.

\subsection{Results from the Gaussian Recovery}
\label{gaussrecoveryresults}

\begin{figure*}
    \centering
    \subfloat{\includegraphics[width=16cm,height=15cm]{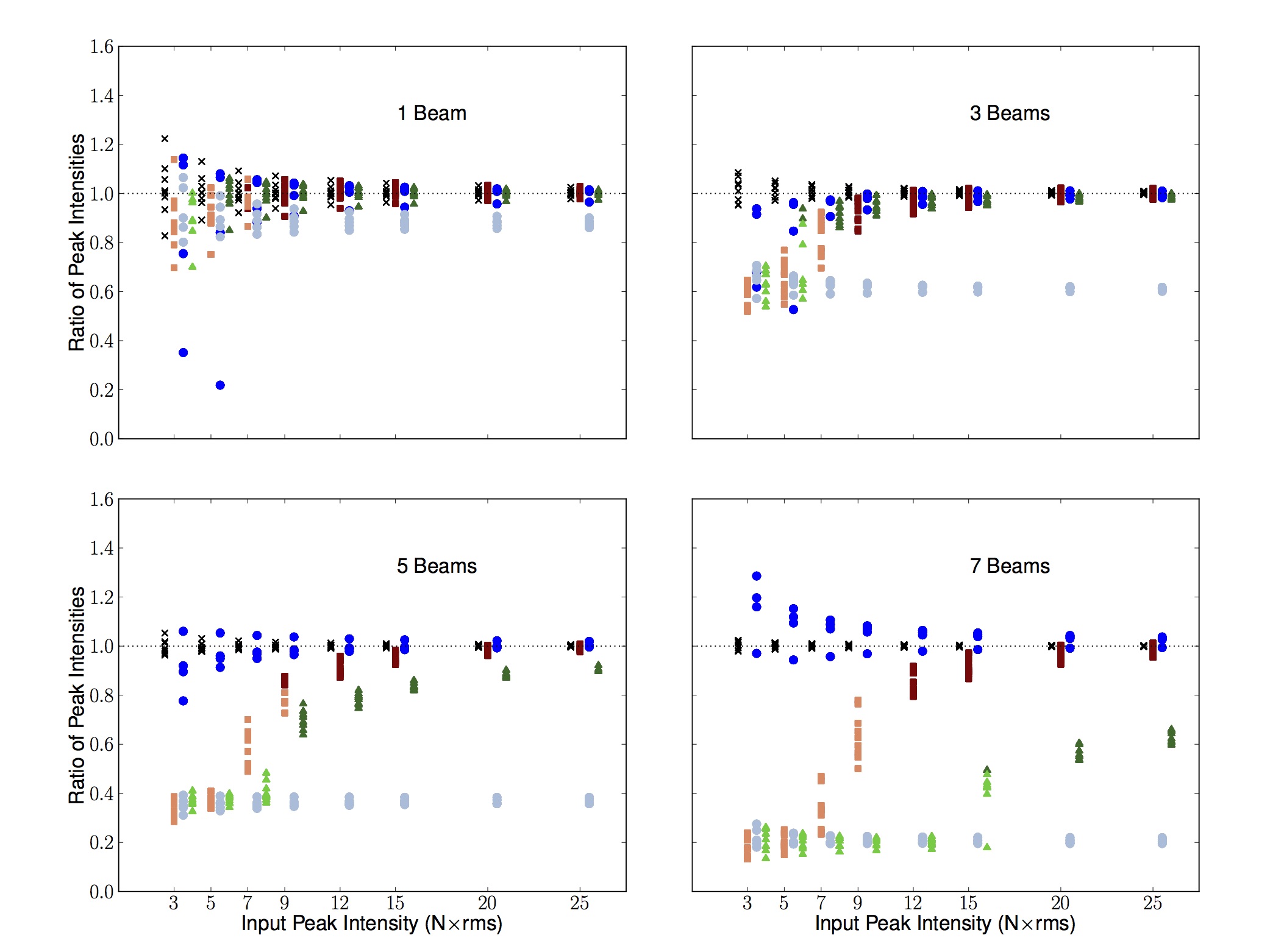}}\\
    \subfloat{\includegraphics[width=16cm,height=3cm]{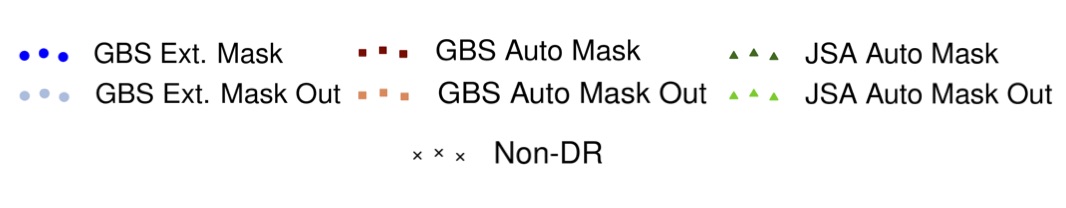}}
     \caption{Artificial source recovery comparison for different data reduction methods: peak intensities. The plot symbols have been separated along the abscissa for better legibility. The ordinate represents the measured output peak intensity divided by the nominal input peak intensity. \emph{Top left}: Gaussians with a 1 beam FWHM. \emph{Top right}: Gaussians with a 3 beam FWHM. \emph{Bottom left}: Gaussians with a 5 beam FWHM. \emph{Bottom right}: Gaussians with a 7 beam FWHM.}
        \label{curvefitpeakfig}
\end{figure*}

\begin{figure*}
\centering
\includegraphics[width=16cm,height=15cm]{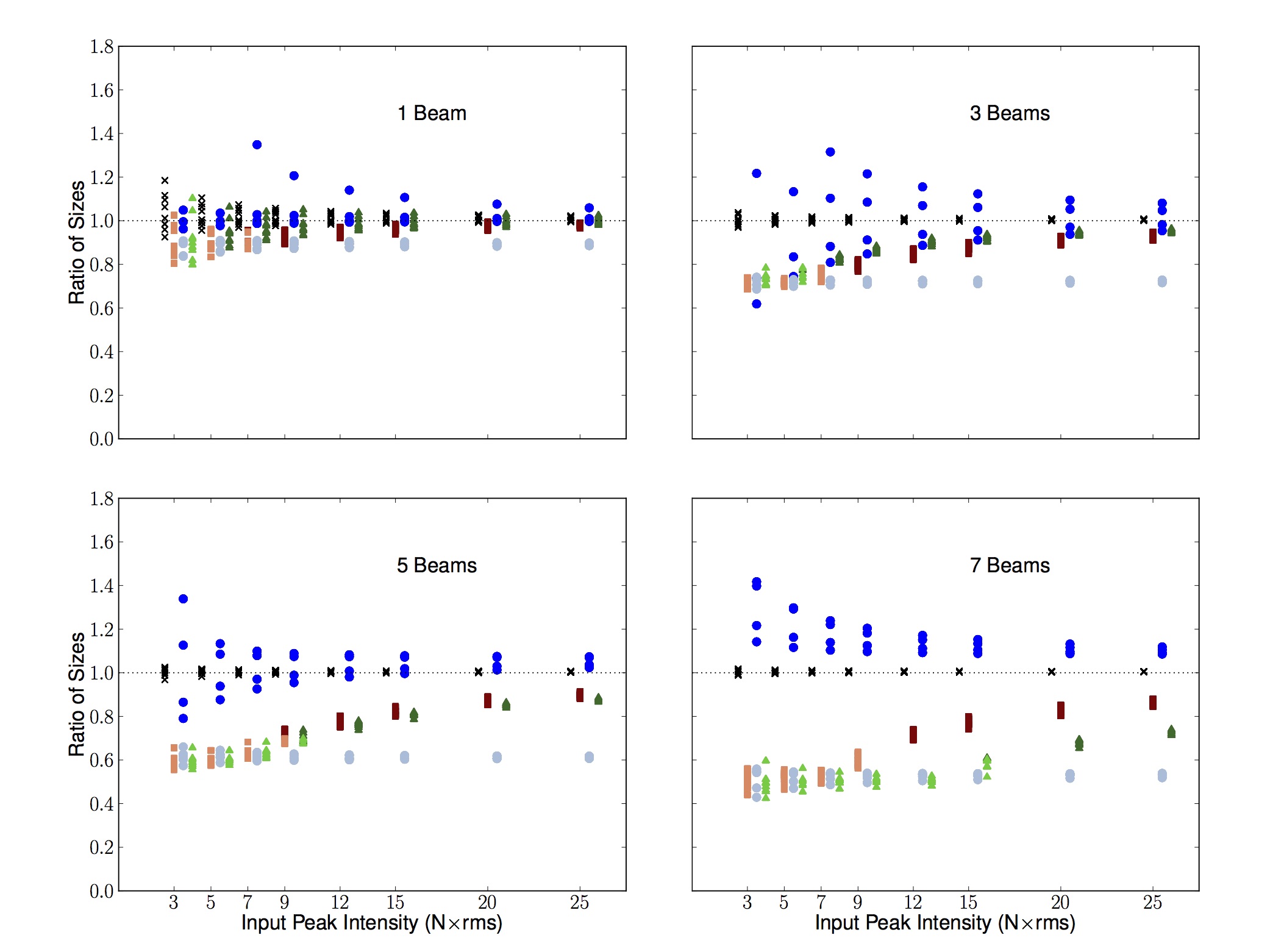}
\caption{Artificial source recovery comparison for different data reduction methods: sizes. The ordinate represents the measured output Gaussian size divided by the nominal input size. The plotting style follows Figure \ref{curvefitpeakfig}.}
\label{curvefitsizefig}
\end{figure*}

\begin{figure*}
\centering
\includegraphics[width=16cm,height=15cm]{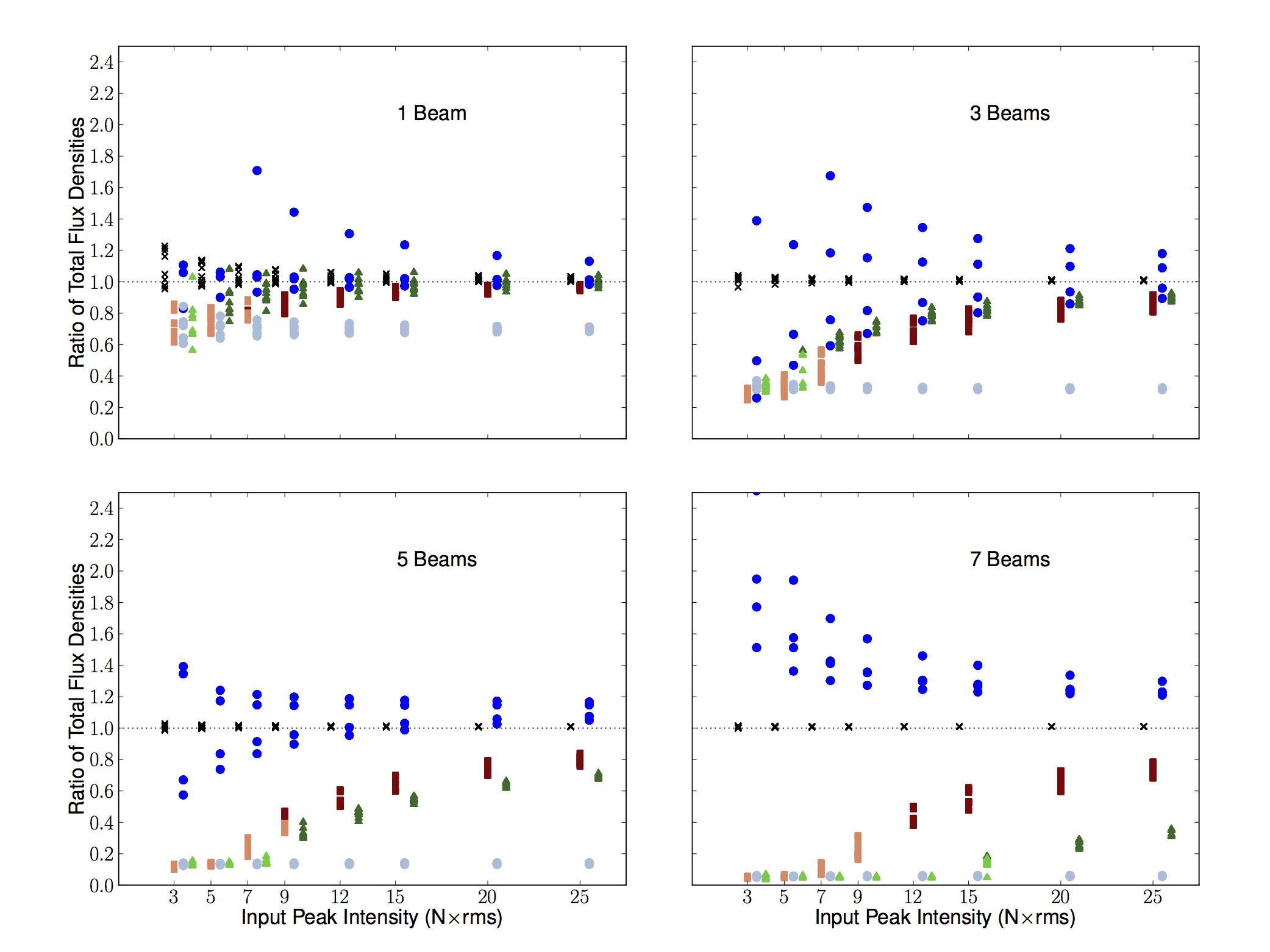}
\caption{Artificial source recovery comparison for different data reduction methods: total flux densities. The ordinate represents the measured output Gaussian total flux density (peak $\times$ size$^{2}$) divided by the nominal input total flux density. The plotting style follows Figure \ref{curvefitpeakfig}.}
\label{curvefittotfluxfig}
\end{figure*}

The results of the Gaussian curve fitter are compared to the nominal Gaussian input values in Figures \ref{curvefitpeakfig} to \ref{curvefittotfluxfig_diffext}. In the following sections, we will address each of the measures used to compare between the output structure and the input objects: peak intensity, size, and total flux density.

In each of the images, only the central nine Gaussians are included in the analysis to avoid the high noise locations on the edge of the image. This was done to mimic the approach to the GBS LR1 data itself as the trusted sources in each map lie far from the noisy edges. 

In Figures \ref{curvefitpeakfig} to \ref{curvefittotfluxfig_diffext}, the ordinate represents the measured Gaussian properties divided by the nominal input properties at each peak intensity for each of the four FWHM values used. There are seven plot symbols used:

\begin{enumerate}[labelindent=0pt,labelwidth=\widthof{\ref{last-item}},label=\arabic*.,itemindent=1em,leftmargin=!]

\item The spatial addition of Gaussians onto the noise field (non-DR) are represented by black Xs. 

\vspace{3mm}

\item Gaussians included in the GBS LR1 external mask (in the images which are produced including an external mask) are represented by dark blue circles.

\vspace{3mm}

\item Gaussians which lie outside the GBS LR1 external mask (in the images which are produced including an external mask) are represented by light blue circles.

\vspace{3mm}

\item Gaussians included in the GBS LR1 automask (in the images which are produced \emph{without} including an external mask) are represented by dark red squares. We define an object to be included in the automask if at least half of the total number of pixels within the respective Gaussian's FWHM are found in the AST automask.

\vspace{3mm}

\item Gaussians which have at least half of the total number of pixels within one FWHM lying outside the GBS LR1 automask (in the images which are produced \emph{without} including an external mask) are represented by light red squares. 

\vspace{3mm}

\item Gaussians included in the JCMT LR1 reduction automask are represented by dark green triangles. Again, we define an object to be included in the automask if at least half of the total number of pixels within the respective Gaussian's FWHM (in all directions) are found in the AST automask.

\vspace{3mm}

\item The Gaussians which lie outside the JCMT LR1 reduction automask are represented by light green triangles.

\end{enumerate}

We performed all reduction tests using the same input Gaussian peak brightness values. In each Figure, however, the symbols have been slightly separated along the abscissa for clarity as they often overlap. 


\subsubsection{Peak Intensity}

In Figure \ref{curvefitpeakfig}, we see that small objects are well recovered even when they are relatively faint. As the Gaussian FWHMs are increased (top left to bottom right), the masked results converge only for brighter Gaussian peaks. For faint objects larger than a point source, however,  the unmasked (light red squares and light blue circles for the GBS LR1, light green triangles for the JCMT LR1 reduction) and automasked (dark red squares for the GBS LR1 reduction, dark green triangles for the JCMT LR1 reduction) cases display significant deviations from the original Gaussian peak values in both reductions. As the Gaussian sizes are increased, the unmasked, reconstructed source properties become lower than those of the source inserted into the map. We also find that as a Gaussian becomes brighter, the given source is recovered more reliably. This behaviour is expected since the pixel-to-pixel variation is steeper for a cut across a Gaussian of a given size with a brighter peak. In contrast, shallow variations tend to get filtered out. Thus, we note that larger Gaussians require brighter peaks for significant pixels to be identified and placed in the automask. 

The GBS LR1 Gaussians within an external mask (dark blue circles) and the non-DR Gaussians (black Xs) display the most consistent results with the input peaks and the GBS LR1 automask reduction (dark red squares) recovering the objects well, even for a 7 beam FWHM source (with a peak of at least 9 $\sigma_{rms}$. Larger objects will need to be even brighter to be detected). There are obvious sources within an external mask that are lying in a negative bowl in the more compact Gaussian grids (see the external masked Gaussians lying significantly below their nominal size in the top right and bottom left panels of Figure \ref{curvefitsizefig}). This negative bowl is part of the artificial structure introduced by an external mask that is too large for the astronomical signal present (see Section \ref{changeexternalmask}). As the sources begin to resemble the mask size, the negative bowl is no longer apparent. For these and smaller sources which are sufficiently bright, the peak intensities are generally accurate to within 10\% of the nominal value, which is similar to the expected calibration uncertainty, while larger and fainter sources are accurate to within 20\% (with some exceptions). The variation seen in the small, faint non-DR Gaussians is simply due to the Gaussian curve fitter algorithm looking for the optimal solution. When it is faint, recovery of a Gaussian source will noticeably suffer more from noise variations, causing the algorithm to find the ``best'' peak. When the structure is small, any deviation from the true location will have greater impact on the peak intensity value and size of the Gaussian fit than the same deviation would have for a larger, smoother source with more pixels near the correct peak value in the same, central vicinity. 

We see also in Figure \ref{curvefitpeakfig} that the JCMT LR1 reduction (triangles) reliably recovers compact structure The peaks found for large objects, however, are significantly underestimated. In both the GBS LR1 and JCMT LR1 reductions, we see that the automasked sources resemble the nominal peak brightness more accurately as the peak brightness is increased before the object is defined to lie within an automask. This behaviour occurs simply because the compact central region of the Gaussian has been identified as it is the brightest location of the source. Less than 50\% of the pixels within one FWHM of the Gaussian centre, however, have been included in the astronomical signal (AST) mask at this point. 

\subsubsection{Size and Total Flux Density}
\label{cfsizesec}

In Figure \ref{curvefitsizefig}, we see results similar to those of the peak intensity for the fitting algorithm's calculated object sizes. It is obvious that the GBS LR1 and JCMT LR1 automask reductions (squares and triangles) miss large structure, as expected. In both cases, however, they identify that at least some structure exists at locations of extended emission, assuming the peak brightness is high enough. The GBS LR1 automask recovers nearly 50\% of the pixels within one FWHM of a 7 beam FWHM Gaussian with a peak of 9 $\sigma_{rms}$, as evidenced by the increase in peak intensities in Figure \ref{curvefitpeakfig}. The JCMT LR1 reduction requires a 15 $\sigma_{rms}$ peak for the same size, however. Once the external mask is applied to this structure in the second step of the GBS LR1 reduction, the values which are measured resemble the original Gaussian properties. Nevertheless, occasional pedestals and bowls are found which can increase or decrease the size of a given source by up to 40\% (see Section \ref{changeexternalmask} for more information).  As discussed above, the GBS LR1 reduction is tuned to recover extended structure whereas the JCMT LR1 reduction is tuned specifically to find compact emission. This means that the JCMT LR1 reduction will always underestimate the large-scale structure present in an observed region.

There are Gaussians that clearly lie on deviations in the noise background induced by the external masking procedure (see Section \ref{changeexternalmask} below for a discussion on external mask size). The noise field used is very uniform (see Figure \ref{drfig}b), however, in Figures \ref{curvefitpeakfig}, \ref{curvefitsizefig}, and \ref{curvefittotfluxfig} we see the recovered compact Gaussians' properties are overestimated and underestimated while the largest Gaussians' properties only exhibit the former simply due to their placement in the map. The masks used here are too large, causing a pedestal effect on some sources and a negative bowl on others. This result reinforces the idea that large-scale noise features present in the map and the size of the external mask around a given source can play an important role in source recovery. Again, just as in the case of the recovered peak intensities, bright sources are found to be within 20\% of their nominal value and many are within 10\%. Looking to the non-DR data (black Xs), we see that there is a fundamental difference in the large-scale structure when we compare with the external masked case (circles). Evidently, in some cases, \emph{makemap} introduces additional structure in the image reconstruction when large external masks are employed and, to conserve flux, it compensates by reducing the real structure present. This behaviour can potentially create bowls around the borders of brighter sources. As the sources increase in size, the external mask's pedestal effect becomes more apparent (see Figures \ref{curvefitsizefig} and \ref{curvefittotfluxfig}). For an example of a bowl, see the 3 beam FWHM and 5 beam FWHM cases in Figure \ref{curvefitsizefig}. The fractional importance of the pedestal or bowl, however, declines with input Gaussian peak strength. Separating the background sky signal from the astronomical signal using a tight boundary around significant structure is, therefore, very important so that \emph{makemap} does not allow features in the noise to grow during subsequent iterations. As we explore below in Section \ref{changeexternalmask}, an external mask should not be larger than twice the size of an emission region if the errors in the measured properties of the source are to remain within 10-20\% of their nominal value (up to the size of the array footprint and/or characteristic high-pass filter scale). This gives the researcher an opportunity to safely mask a whole region with a generous boundary in case there is indeed faint extended structure that is not apparent from a simple automask reduction.
	
It is important to reiterate that in the GBS LR1 reduction, the external mask size which is used is based on the previously performed automask reduction (see Section \ref{GBSreductionsec}). In this automask reduction, the common mode subtraction over the spatial scale of the bolometer array at each time step and the filter size acting on the time stream are the most important considerations when determining the scales we can trust in the final images (see parameters \texttt{com.perarray} and \texttt{flt.filt\_edge\_largescale} in Section \ref{dimmconfigsec}; also see \citealt{chapin2013}). The full SCUBA-2 focal plane, including all four subarrays, is $400\arcsec \times 400\arcsec$. This nominally means that objects with sizes up to $\sim$400$\arcsec$ can be confidently recovered during the common mode subtraction at each time step. Large objects approaching this scale, however, can create a similar signal in a high percentage of bolometers causing them to be targeted as common mode, low frequency noise. Thus, large-scale modes can be subtracted from these structures, diminishing their sizes and leading to uncertainties in drawing the external mask boundaries.

In addition to this common mode subtraction, the filter operating on the time stream was chosen to be 600$\arcsec$, approximately the diagonal size of the full 400$\arcsec$$\times$400$\arcsec$ focal plane array. As noted in Section \ref{dimmconfigsec}, this filter uses the scanning speed of the JCMT combined with the provided spatial scale to subtract large-scale modes from each individual bolometer time series. Thus, on scales of $\sim$400$\arcsec$ to 600$\arcsec$ and larger (depending on the external mask boundaries), significant structures are recovered but with diminished sizes and total fluxes due to the subtraction of large-scale modes (see Section 2 of \citealt{pattle2015}). The degeneracy between significant large-scale sources and the common mode, however, can cause artificial structure to arise in various parts of the map (see \citealt{chapin2013}, section 4). Therefore, drawing external masks which are larger than necessary can fuel these degeneracies and create obvious pedestals and bowls. Note that in the case of the JCMT LR1 reduction, we cannot reliably identify structures larger than 200$\arcsec$ as this is the time stream filter scale as well as the subarray footprint (the common mode is calculated over each individual subarray in this reduction) and no external mask reduction is subsequently performed.			
		
An interesting question for non-artificial sources is how much real structure we are missing at large-scales. An automask will certainly pick up the brightest inner locations of the larger Gaussians, but, the correct size of the external mask to use surrounding this area is clearly debatable. An external mask which is too small will miss valuable structure but one that is too large will result in artificial structure, as mentioned above (also see Section \ref{changeexternalmask} for a more quantitative analysis). Large Gaussians which are not included in the external mask will evidently be missed, especially faint objects, and there is already an appreciable amount of unrecovered emission ($\sim$30\%) missing from the unmasked and automasked Gaussians with FWHMs of only 3 beams (see the squares, triangles, and light blue circles).


\section{Other Reduction Considerations}
\label{changemasks}

\subsection{Changing the Automask Parameters}
\label{changeautomask}

To explore how an automasked reduction would differ using different thresholds for the AST model, we change the parameters \emph{ast.zero\_snr} (the threshold at which significant structure is identified) and the \emph{ast.zero\_snrlo} (the flux level of the surrounding pixels to which the identified significant peaks will be extended) using the GBS LR1 automask reduction method (see Table \ref{automasktable}). Since the automask reduction accurately recovers compact structures over a broad range of these two parameters and borderline island detections are found within the extended emission, we only perform this analysis for 7 beam (105\arcsec) FWHM Gaussians. Note that setting the \emph{ast.zero\_snrlo} parameter to 0 does not allow any extension to flux levels lower than the threshold defined by \emph{ast.zero\_snr}.

\begin{table}
\caption{Summary of the \emph{ast.zero\_snr} and \emph{ast.zero\_snrlo} parameters tested. The \emph{ast.zero\_snr} parameter represents flux threshold for identifying astronomical signal. The \emph{ast.zero\_snrlo} parameter allows (or disallows if it is set to 0) identified sources with pixel values of at least the flux threshold defined by \emph{ast.zero\_snr} to expand in area until a second flux threshold is met. Bold font indicates the current GBS LR1 automasking parameters investigated in Section \ref{completenesssec}.}
\label{automasktable}
\begin{tabular}{|c|c|c|}
\hline
\emph{ast.zero\_snr} & \emph{ast.zero\_snrlo} & Gaussian FWHM\\
\hline\hline
\textbf{5} &  \textbf{0} & \textbf{7 beams}  \\
5 & 3 &  7 beams  \\
5 & 2 & 7 beams \\
3 & 2 & 7 beams\\
\hline
\end{tabular}
\end{table}

\begin{figure*}
\centering
\includegraphics[width=16cm,height=15cm]{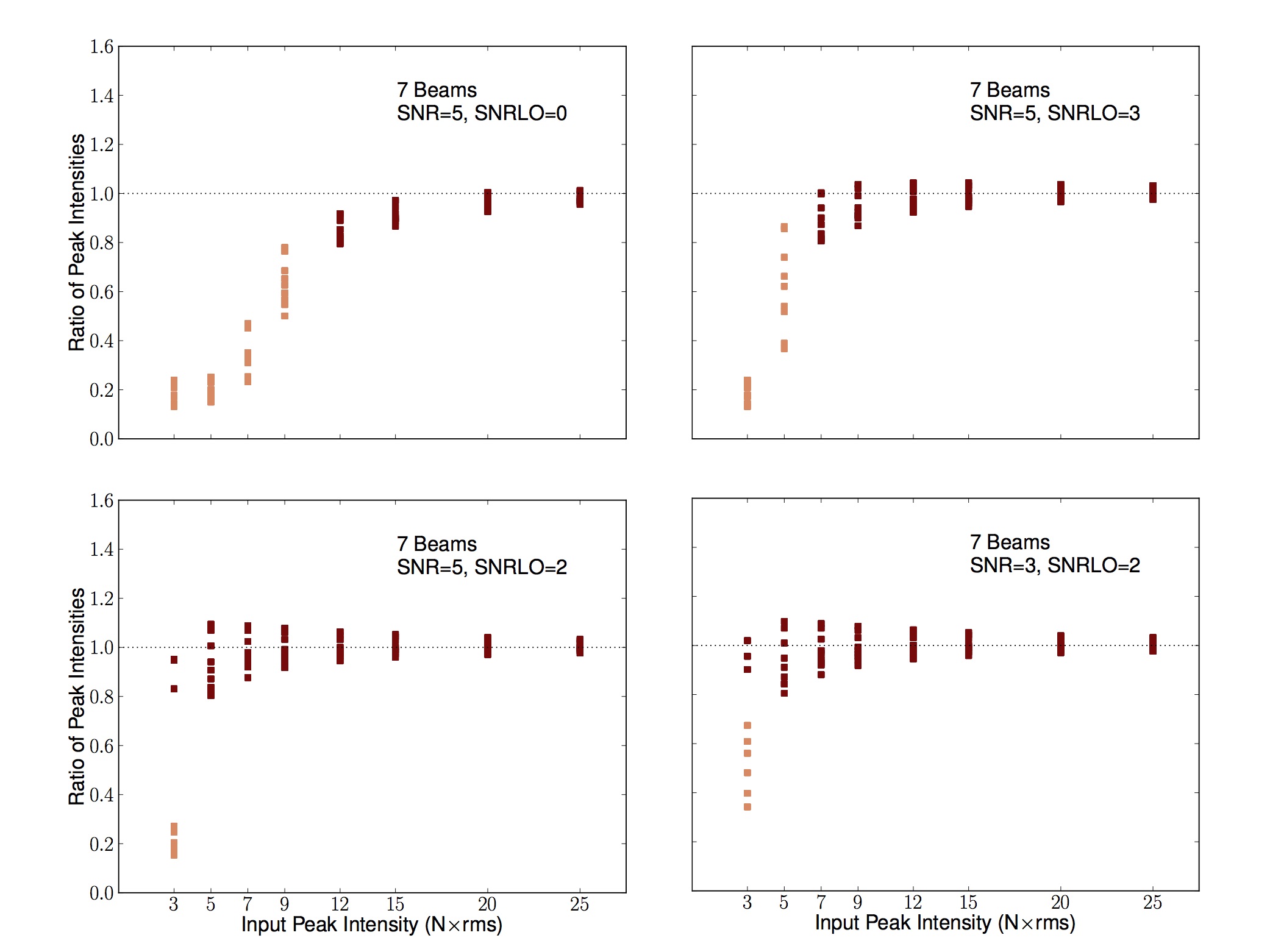}
\caption{Artificial source recovery comparison for different GBS LR1 automask parameters: peak intensities. The ordinate represents the measured output peak intensity divided by the nominal input peak intensity. Light red indicates that the object had less than 50\% of the pixels within one FWHM of the peak location detected in the AST mask, dark red indicates it had at least 50\%. \emph{Top left}: \emph{ast.zero\_snr} = 5, \emph{ast.zero\_snrlo} = 0, the original GBS LR1 automask parameters.  \emph{Top right}: \emph{ast.zero\_snr} = 5, \emph{ast.zero\_snrlo} = 3. \emph{Bottom left}: \emph{ast.zero\_snr} = 5, \emph{ast.zero\_snrlo} = 2. \emph{Bottom right}: \emph{ast.zero\_snr} = 3, \emph{ast.zero\_snrlo} = 2.}
\label{curvefitpeakfig_diffauto}
\end{figure*}


\begin{figure*}
\centering
\includegraphics[width=16cm,height=15cm]{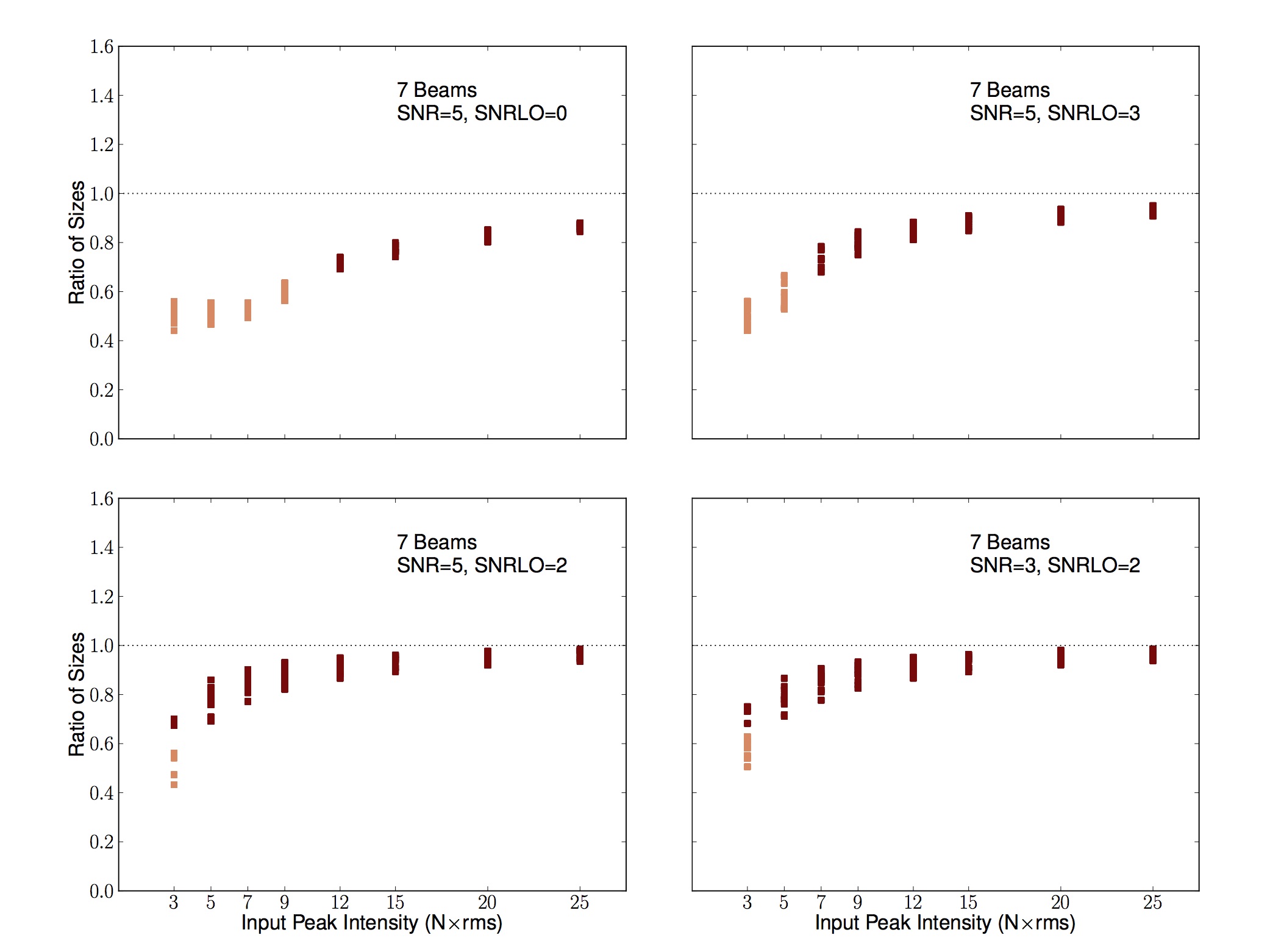}
\caption{Artificial source recovery comparison for different GBS LR1 automask parameters: sizes. The ordinate represents the measured output size divided by the nominal input size. The plotting style follows Figure \ref{curvefitpeakfig_diffauto}.}
\label{curvefitsigmafig_diffauto}
\end{figure*}


\begin{figure*}
\centering
\includegraphics[width=16cm,height=15cm]{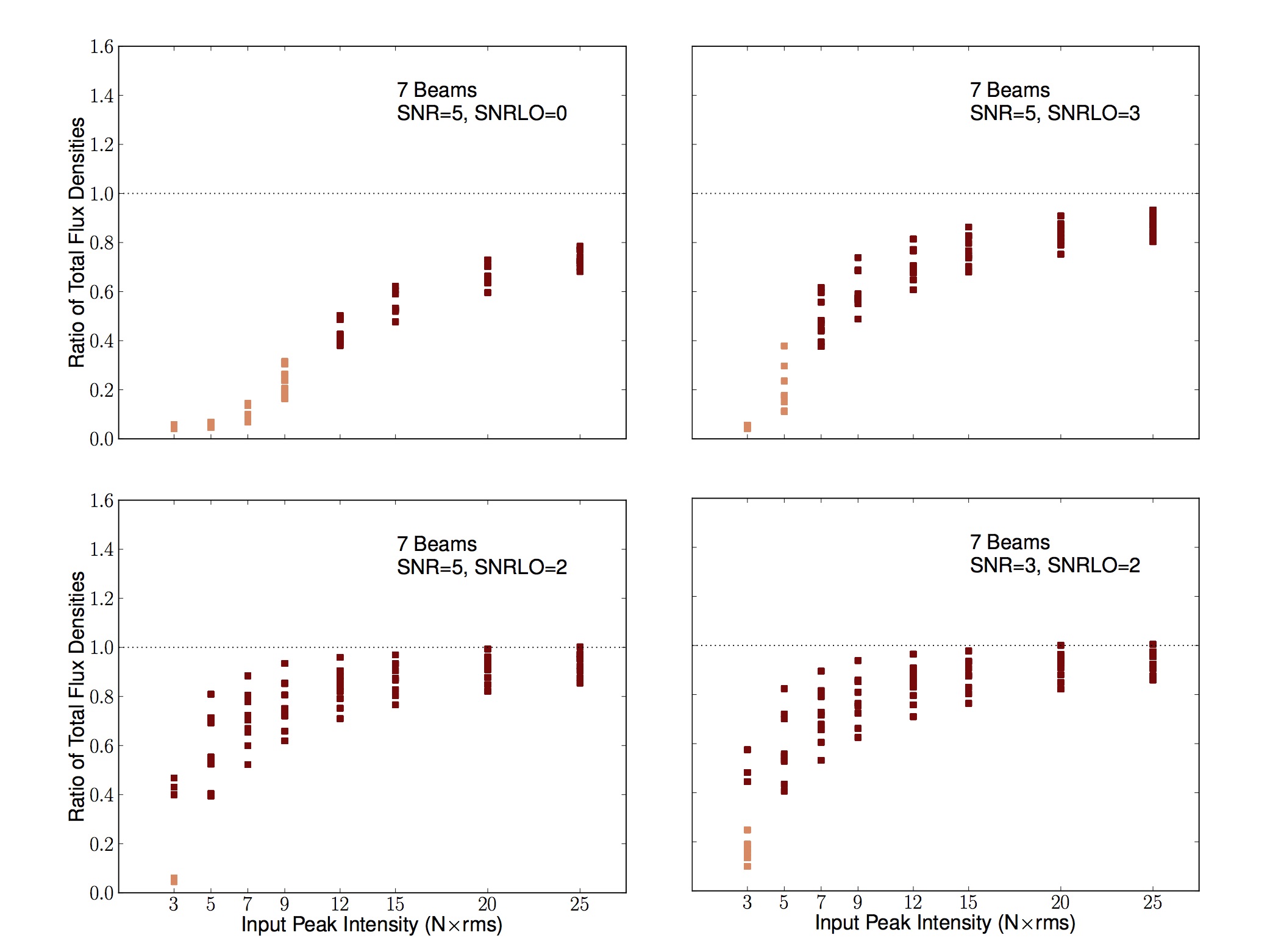}
\caption{Artificial source recovery comparison for  different GBS LR1 automask parameters: total flux densities. The ordinate represents the measured output total flux density divided by the nominal input total flux density. The plotting style follows Figure \ref{curvefitpeakfig_diffauto}.}
\label{curvefittotfluxfig_diffauto}
\end{figure*}

\subsubsection{Peak Intensity}

In Figure \ref{curvefitpeakfig_diffauto}, we compare four different automask reductions. The reduction using the original GBS LR1 automask parameters are shown in the top left and the other three reductions were performed with the \emph{ast.zero\_snr} and \emph{ast.zero\_snrlo} parameters in the \emph{dimmconfig} file changed to the values shown. Clearly, when the \emph{ast.zero\_snrlo} parameter is not used, significant structure in faint sources is missed in the AST automask. When one allows identified structures with masked pixels (brightnesses above 5 $\sigma_{rms}$) to grow down to a level of 2 $\sigma_{rms}$ or 3 $\sigma_{rms}$, however, much more of the expected Gaussian brightness is recovered, especially for the fainter input peaks. This improvement happens because when the constraint on the minimum flux included in the AST model is relaxed, more of the full Gaussian extent is identified as astronomical signal earlier in the iterative map making process, allowing more of the significant emission to be extracted from the noise. 

\subsubsection{Size and Total Flux Density}

A similar trend is seen in the recovered sizes and total flux densities (Figures \ref{curvefitsigmafig_diffauto} and \ref{curvefittotfluxfig_diffauto}) as in the recovered peak brightnesses, above. When the \emph{ast.zero\_snrlo} parameter is set to zero, much of the expected emission lies outside the AST mask for fainter objects. When the automask is extended from the identified 5 $\sigma_{rms}$ peaks down to a level of 2 $\sigma_{rms}$ or 3 $\sigma_{rms}$, we see that more significant emission is recovered in the AST mask. As expected, the measured total flux density shows very similar results to the size of the Gaussian structures recovered by the automask. 

\subsection{External Mask Size}
\label{changeexternalmask}

Due to the observed pedestals and bowls induced by the size of a masked region around a source, we perform a further test of the external mask reductions where we change the size of each mask in the checkerboard (see Figure \ref{extmaskfig}). Table \ref{extmasktable} outlines the different mask sizes explored, with bold font indicating the original external mask size used in the analysis presented in Section \ref{completenesssec}. Again, tests are performed only for the 7 beam FWHM Gaussians using the GBS LR1 external mask data reduction and in this case we only observed a pedestal effect (the bowls arose from different Gaussian grid configurations).

\begin{table}
\caption{Summary of the sizes of the square masks in the checkerboard style external mask tests. The ``size'' indicated here is the length of the sides of the square external masks placed over every second Gaussian. Bold font indicates the original external mask size investigated in Section \ref{completenesssec}.}
\label{extmasktable}
\begin{tabular}{|c|c|}
\hline
Size of Square Patch in Checkerboard & Gaussian FWHM\\
\hline\hline
4 beams & 7 beams \\
12 beams & 7 beams \\
\textbf{20 beams} &  \textbf{7 beams}  \\
36 beams & 7 beams \\
\hline
\end{tabular}
\end{table}

\begin{figure*}
\centering
\includegraphics[width=16cm,height=15cm]{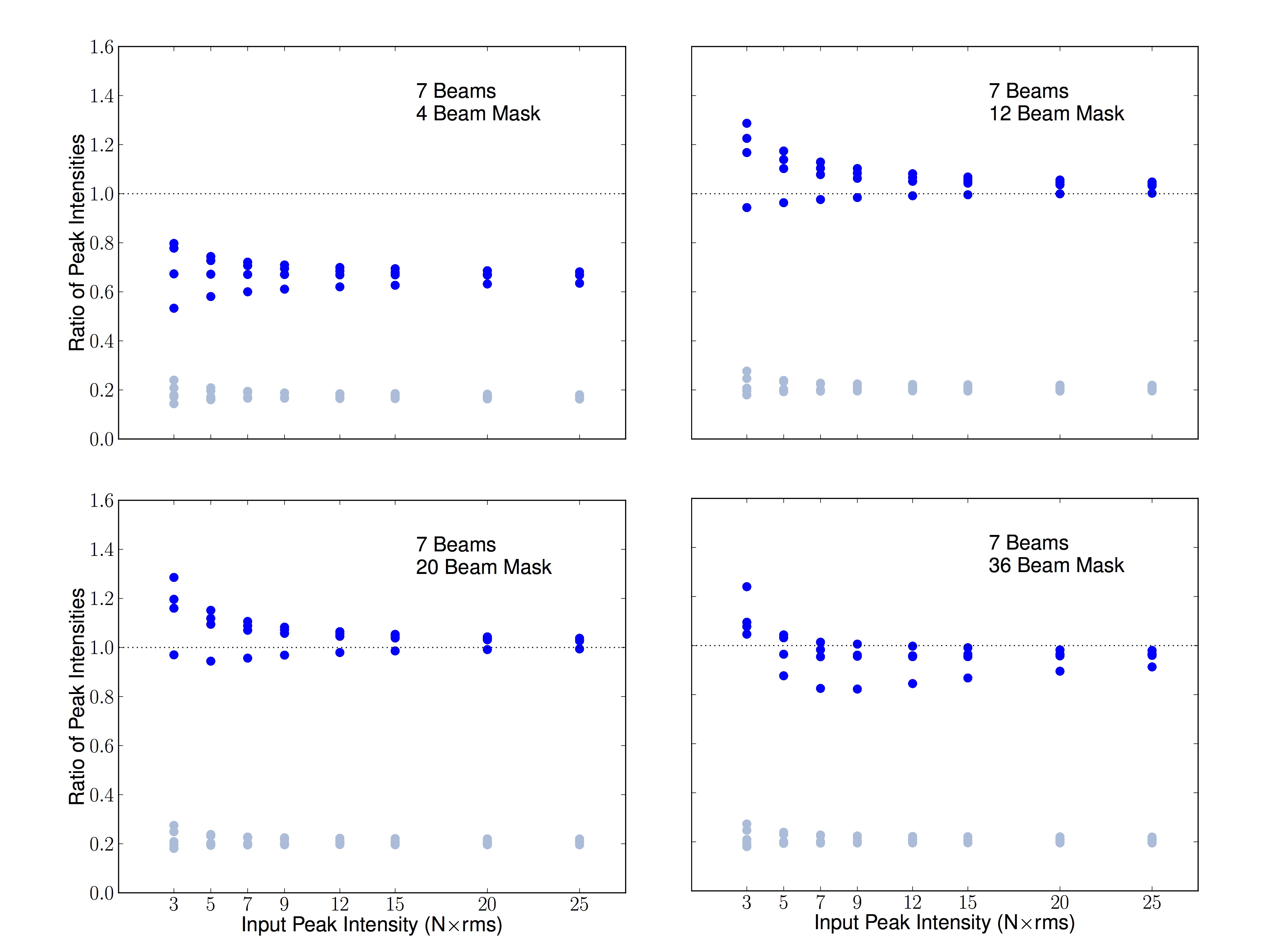}
\caption{Artificial source recovery comparison for different external mask sizes: peak intensities. The ordinate represents the measured output peak intensity divided by the nominal input peak intensity. Light blue indicates that the object was outside the mask, dark blue indicates it was inside the mask. \emph{Top left}: 4 beam masks.  \emph{Top right}: 12 beam masks. \emph{Bottom left}: 20 beam masks (original). \emph{Bottom right}: 36 beam masks.}
\label{curvefitpeakfig_diffext}
\end{figure*}


\begin{figure*}
\centering
\includegraphics[width=16cm,height=15cm]{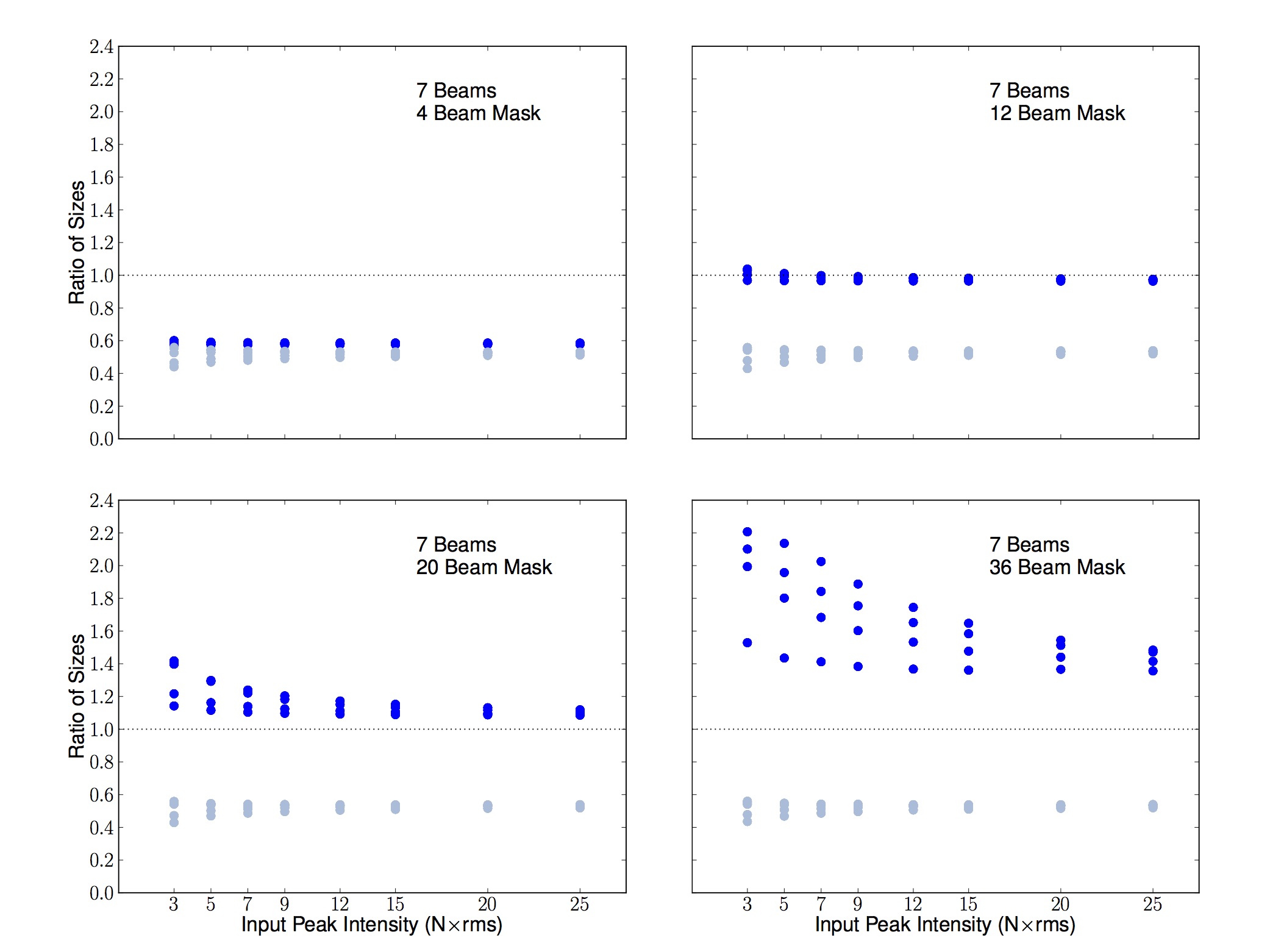}
\caption{Artificial source recovery comparison for  different external mask sizes: sizes. The ordinate represents the measured output size divided by the nominal input size; note the change in the ordinate's range from the figures above so the data points would be visible on all panels. The plotting style follows Figure \ref{curvefitpeakfig_diffext}.}
\label{curvefitsigmafig_diffext}
\end{figure*}


\begin{figure*}
\centering
\includegraphics[width=16cm,height=15cm]{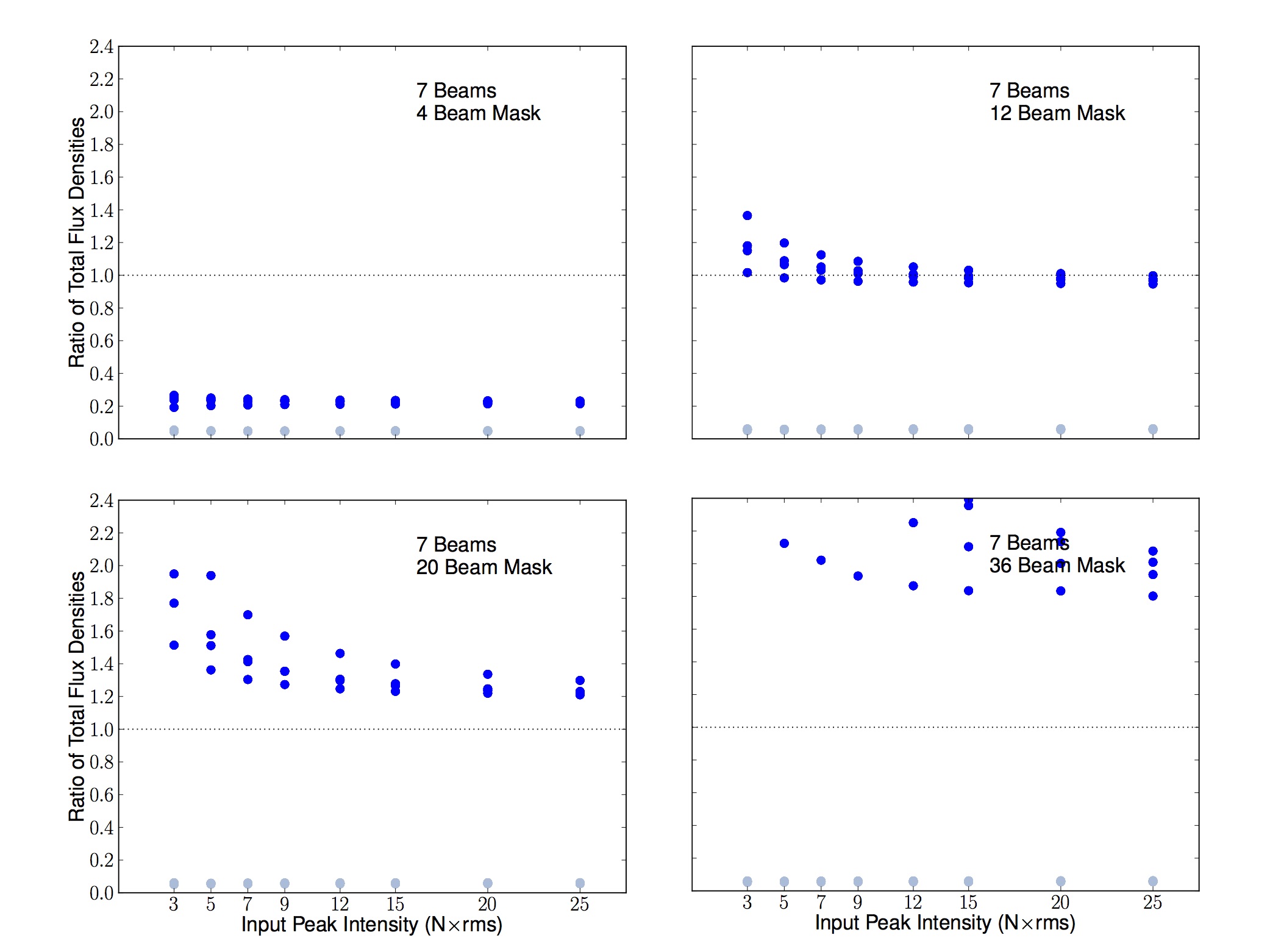}
\caption{Artificial source recovery comparison for  different external mask  sizes: total flux densities. The ordinate represents the measured output total flux density divided by the nominal input total flux density; note the change in the ordinate's range from the figures above so the data points would be visible on all panels. The plotting style follows Figure \ref{curvefitpeakfig_diffext}.}
\label{curvefittotfluxfig_diffext}
\end{figure*}

\subsubsection{Peak Intensity}

In Figure \ref{curvefitpeakfig_diffext}, we see that the application of different sizes of external masks can play a key role in the extracted data. For the small, 4 beam mask cases, much of the flux is suppressed since the Gaussian FWHM itself is almost double the mask size. Once the mask is larger than the Gaussian, however, we see that the expected peak intensity of the object is returned reasonably well, even in the faintest cases. For masks larger than 4 beams, we see the introduction of artificial structure (up to $\sim$30\%) and a higher uncertainty for the fainter sources as the pedestal effect increases for larger masks. Brighter peaks in general result in more accurate measurements, as expected, since more of the Gaussian emission will be more significant relative to the noise.

\subsubsection{Size and Total Flux Density}

Again, in the size and total flux density plots (Figures \ref{curvefitsigmafig_diffext} and \ref{curvefittotfluxfig_diffext}) we see similar results to the peak intensities discussed above. In the smallest 4 beam mask case, we see tightly constrained recoveries that miss the outer emission. This result is expected because we are focusing on the brightest region of each Gaussian. A tighter mask that is large enough to cover the whole source (e.g., 12 beams in our case) produces the best results. It is clear that uncertainties grow rapidly for the faint sources. Also, in the 36 beam mask case, the brighter objects are also affected by artificial large-scale structure introduced by \emph{makemap}. Ideally, the mask should just encompass the source with little background flux at the edges to ensure none of the fainter extended emission is missed and that no substantial noise is included in the external mask. By using an external masking procedure based on a rigorous analysis for each individual region observed by the GBS, it is unlikely that the 850 $\mu$m maps produced for the GBS LR1 reduction will suffer from mask areas overestimated by much greater than twice the size of the source. Thus, the reduction performed with the 12 beam external mask masks in this section shows the closest example to the real GBS LR1 data. The measured total flux density shows very similar behaviour to the size plots, as expected.




\section{Data Reduction and Common Physical Measurements}
\label{observablessec}

Evidently, the JCMT LR1 and GBS LR1 data reduction methods recover different amounts of extended structure. In this section, we qualitatively discuss the impact of using these different data reduction techniques on two common physical measurements in star-forming regions: 1. the core mass function (CMF) and 2. the derivation of the temperature. Note that in any project, the reduction method should be chosen based upon the specific scientific goals one is researching with an understanding of the benefits as well as the drawbacks offered by that method.

\subsection{The Core Mass Function}
\label{CMF}

There have been many studies on the mass distribution of core populations derived from the flux due to dust in nearby star forming regions (for examples, see \citealt{johnstone2000}, \citealt{wardthompsonPPV}, \citealt{enoch2008}, and \citealt{sadavoy2010}). Although this measurement is intrinsically dependent on the core identification algorithm used, specifically in the ways that boundaries are drawn around cores, the type of data reduction employed can also inform the final results. Along with the definition of the AST mask, the main reduction parameters under consideration when analysing the mass of extended objects are those which govern the time stream filtering and the common mode low-frequency noise removal.

In the optically thin ($\tau << 1$), isothermal limit, the dust emission traces the mass of a given object. A generally accepted result is that the more massive objects are not just brighter but also larger (see, for example, \citealt{sadavoy2010}'s Figure 12). With harsh filtering criteria like those employed in the JCMT LR1 reduction, the most extended structure is not recovered and the AST mask consists of fairly tight boundaries around the most concentrated, compact regions of the map. The common mode noise is also estimated over each individual subarray, limiting the size of confidently detected structure to 200$\arcsec$. This may not result in an accurate assessment of the population of core masses as any larger objects will be ``missing'' flux. Thus, since larger objects correlate with larger intrinsic masses, the core mass function may be steepened. 


The GBS LR1 is a much more robust reduction technique to use for determining core mass functions when compared to the JCMT LR1 assuming the same core identification algorithm. The larger size over which the common mode is estimated, the more relaxed filtering parameters, and the greater number of iterations performed work to recover much more significant structure than just the compact sources. The two-stage masking process is also highly beneficial for measurements like the core mass distribution. This ensures that two passes are taken at recovering the maximum amount of structure, including a robust analysis that is not automated. To improve this even further, we have suggested that the next generation of the GBS data reduction be performed with different \textit{ast.zero\_snr} and \textit{ast.zero\_snrlo} parameters to potentially recover more faint structure (see Sections \ref{changeautomask} and \ref{conclusionsec}). Of course, as previously discussed, large-scale modes will still be removed from the final map so it is very important to ensure that this is taken into account. Comparing the SCUBA-2 data with Herschel Space Observatory data is one method that can be used for better understanding the large-scale structure present in a given region.

\subsection{Derivation of the Temperature with SCUBA-2}

Another common measurement performed with SCUBA-2 data involves the temperature maps produced by calculating the 850$\mu$m/450$\mu$m intensity ratio, or against other data sets (such as the Herschel Space Observatory). 
In order to perform this measurement, the beams should be matched between the two data sets and the filtering parameters as well as the masks should be identical. The large-scale noise present in each individual map, however, will differ and this uncertainty becomes very important when comparing two maps.
Preliminary work performed by GBS team members (Hatchell \& Rumble, private communication; also see \citealt{hatchell2013}) based on the GBS LR1 reduction suggests that including structures larger than 300$\arcsec$ indeed leads to unrealistic ratios between 850$\mu$m and 450$\mu$m. In order to perform confident measurements, that data must be post-processed with a tighter filter so that only the inner regions of these large objects have accurate associated temperatures.

When determining the extent of the trustworthy temperature calculation, the profile of the structure itself must also be taken into consideration. It will be very difficult to measure accurate ratios on large, faint emission due to the subtraction of large-scale modes. Again, because of the harsh filtering parameter and the common mode removal (see Section \ref{CMF}, above), the JCMT LR1 reduction method would be appropriate for studying the temperatures of bright, compact objects. Calculating a temperature map for any emission that lies a significant distance from peaked structure relative to the sub-array footprint (< 200$\arcsec$) would be very uncertain. In addition, it is important to consider that the signal to noise ratio is typically lower in the SCUBA-2 450$\mu$m data set (compared to the 850$\mu$m data) so artificial structure is likely to be more prominent in these observations.

\section{Conclusions}
\label{conclusionsec}

In this paper, we have presented and compared two methods of data reduction: the JCMT LR1 reduction (Graves et al. in prep.) and the GBS LR1 reduction (Kirk et al. in prep.) which both employ the \emph{makemap} algorithm \citep{chapin2013} executed with different configurable parameters. We used data from the Orion A South star forming region (Mairs et al. in prep.) to characterise the differences in the scale of the emission and the source morphology recovered by the two methods. We have measured source peak intensities, total flux densities, and radii across three representative regions for each reduction method and compared the results both qualitatively and quantitatively. To further our investigation, we created a series of artificial Gaussians (varied in size and peak brightness) and inserted them into the time domain of a pure noise field using each data reduction method. We then recovered each source using a Gaussian fitting algorithm and compared the measured properties to the nominal input values to observe how much emission was preserved at each scale. Note that although inserting Gaussians into the time stream is an effective way to gain insight into the data reduction process, real structures present in the GBS maps are non-Gaussian and in fact most regions display complex multi-scale structure. To summarise our main conclusions, we find: 

\begin{enumerate}[labelindent=0pt,labelwidth=\widthof{\ref{last-item}},label=\arabic*.,itemindent=1em,leftmargin=!]

\item Both reductions recover the peak intensities of bright compact sources (sources with FWHMs of 3 beams or less and a peak brightness of at least 7 $\sigma_{rms}$) to within 10-20\% of the true value. The GBS LR1 reduction also accurately recovers the peak intensities of the larger Gaussians whereas the peaks produced by the JCMT LR1 reduction are diminished because of the stringent filtering parameters. As expected, there is more uncertainty in the size and total flux density measurements for both reductions (see below). In general, we see more accurate results for objects which are both brighter and more compact. 

\vspace{3mm}

\item Although the JCMT LR1 reduction only recovers compact emission present in a given region and the GBS LR1 reduction recovers the extended structure, the two reduction methods trace the same general areas of significant signal very closely. The JCMT LR1 reduction's diminished extended structure, however, causes island identification algorithms to break up large areas into many smaller objects. Conversely, the GBS LR1 reduction draws out these locations of extended emission and, thus, much more emission can be recovered. These differences are illustrated in Figures \ref{orionasouthcompareregion0} to \ref{orionasouthcompareregion2}.

\vspace{3mm}


\item For faint objects larger than a point source, only a fraction of the true size (and, therefore, total flux density) originally present in an artificial Gaussian is recovered without an external masking procedure. The larger the object, the brighter it must be for an automask detection. In the GBS LR1 automask data reduction, Gaussians with a FWHM of 7 beams must have a peak brightness of 5 $\sigma_{rms}$ for the peak intensity to be measured to within  $\sim$20\%. To measure the total flux density to within 20\% of the nominal value, a peak brightness of greater than 25 $\sigma_{rms}$ is required. For a Gaussian with a FWHM of 7 beams, the JCMT LR1 reduction will never result in \textit{jsa\_catalogue} being able to measure a source's peak intensity or total flux density to within 20\% accuracy of the nominal value due to the inherent filtering of large spatial scales. See Section \ref{gaussrecoveryresults} and Figures \ref{curvefitpeakfig} through \ref{curvefittotfluxfig_diffext}.  

\vspace{3mm}

\item When identifying objects with the automask employed in the GBS LR1 automask reduction, the most accurate Gaussian parameters are measured when nearby pixels are incorporated by extending to lower flux values. For improved reductions, e.g. GBS Legacy Release 2 (GBS LR2), we recommend setting the \emph{ast.zero\_snrlo} parameter to a value of 2. Similarly, the \emph{ast.zero\_snr} parameter should be set to 3. See Section \ref{changeautomask} for details. One must, however, be cautious when extending the AST mask around significant peaks down to a lower flux threshold as artificial structure could be introduced around noise spikes. Fortunately, in the variety of automask tests performed, there was no significant evidence of any artificial structure outside of the Gaussians being included in the AST model.

\vspace{3mm}

\item The external mask used in the GBS LR1 reduction can increase or decrease the brightness of compact peaks depending on the surrounding region. Thus, in some cases where an incorrect external mask size is used, the JCMT LR1 reduction and the GBS LR1 reduction can even differ a little on compact scales (see Figures \ref{orionasouthcompareregion0} to \ref{orionasouthscatterplot}). Artificial structure in the data caused by a poorly sized external mask can act as pedestals and bowls, affecting the peak intensity, the size, and the total flux density measured for a given object. A mask that is too small will only highlight compact regions embedded within larger extended emission. A mask that is too large will include noise variations in the AST mask. A reasonable external mask should cover a given source in its entirety and extend a small distance into the noise. It should, however, be less than a factor of 2 larger than a source which was reliably recovered by the automask reduction in order to achieve 20\% accuracy in total flux density (see Section \ref{cfsizesec} for a discussion on trustworthy scales). In Section \ref{changeexternalmask}, we find that this is possible even for faint, large sources.

\end{enumerate}

\section*{Acknowledgements}

We would like to thank the anonymous
referee for improving this manuscript with their constructive comments.
Steve Mairs was partially supported by the Natural Sciences and
Engineering Research Council (NSERC) of Canada graduate scholarship
program. Doug Johnstone is supported by the National Research Council
of Canada and by an NSERC Discovery Grant. The James Clerk Maxwell Telescope has historically been operated by the Joint Astronomy Centre on behalf of the Science and Technology Facilities Council of the United Kingdom, the National Research Council of Canada and the Netherlands Organisation for Scientific Research. Additional funds for the construction of SCUBA-2 were provided by the Canada Foundation for Innovation. This research has made use of NASA's Astrophysics Data System and the facilities of the Canadian Astronomy Data Centre operated by the National Research Council of Canada with the support of the Canadian Space Agency.
	
\bibliography{dr_apj_mairs}

\end{document}